\documentclass[a4paper,11pt]{article}
\usepackage{jinstpub} 
\usepackage{lineno}
\usepackage{subfig}
\usepackage{placeins}

\title{\boldmath Laser calibration of the ATLAS Tile Calorimeter during LHC Run~2}

\author[a]{M. N. Agaras}
\author[b]{A. Ahmad}
\author[c]{A. Blanco}
\author[d]{D. Boumediene}
\author[d]{R. Bonnefoy}
\author[d]{D. Calvet}
\author[e]{M. Calvetti}
\author[d]{R. Chadelas}
\author[c,f]{P. Conde Muino}
\author[g]{A. Cortes~Gonzalez}
\author[d]{M. Crouau}
\author[d]{C. Crozatier}
\author[d]{F. Daudon}
\author[h]{T. Davidek}
\author[i]{G. Di~Gregorio}
\author[j]{L. Fiorini}
\author[c]{B. Galhardo}
\author[d]{Ph. Gris}
\author[b]{P. Klimek}
\author[d]{P. Lafarguette}
\author[d]{D. Lambert}
\author[e]{S. Leone}
\author[c]{A. Maio}
\author[k]{M. Marjanovic}
\author[c]{F. Martins}
\author[b]{M. Mlynarikova}
\author[c]{B. Pereira}
\author[c,1]{R. Pedro,\note{Corresponding author.}}
\author[h]{K. Petukhova}
\author[h]{S. Polacek}
\author[l]{R. Rosten}
\author[d]{C. Santoni}
\author[e]{F. Scuri}
\author[d]{D. Simon}
\author[m]{Y. Smirnov}
\author[n]{A. Solodkov}
\author[d]{O. Solovyanov}
\author[b]{M. Van Woerden}
\author[c]{F. Veloso}
\author[b]{H. Wilkens}

\affiliation[a]{Institut de F\'{i}sica d’Altes Energies (IFAE), Barcelona Institute of Science and Technology, Barcelona; Spain}
\affiliation[b]{CERN, Geneva; Switzerland}
\affiliation[c]{Laborat\'{o}rio de Instrumenta\c{c}\~{a}o e F\'{i}sica Experimental de Part\'{i}culas - LIP, Lisboa; Portugal}
\affiliation[d]{LPC, Universit\'{e} Clermont Auvergne, CNRS/IN2P3, Clermont-Ferrand; France}
\affiliation[e]{INFN Sezione di Pisa; Italy}
\affiliation[f]{Departamento de F\'{i}sica, Instituto Superior T\'{e}cnico, Universidade de Lisboa; Portugal}
\affiliation[g]{Institut für Physik, Humboldt Universität zu Berlin, Berlin; Germany}
\affiliation[h]{Charles University, Faculty of Mathematics and Physics, Prague; Czech Republic}
\affiliation[i]{APC, Universit\'{e} Paris Cit\'{e}, CNRS/IN2P3, Paris; France}
\affiliation[j]{Instituto de F\'{i}sica Corpuscular (IFIC), Centro Mixto Universidad de Valencia - CSIC, Valencia; Spain}
\affiliation[k]{Homer L. Dodge Department of Physics and Astronomy, University of Oklahoma, Norman OK; United States of America}
\affiliation[l]{Ohio State University, Columbus OH; United States of America}
\affiliation[m]{Northern Illinois University, DeKalb; United States of America}
\affiliation[n]{Institute for High Energy Physics of the National Research Centre Kurchatov Institute, Protvino; Russia}

\emailAdd{rute.pedro@cern.ch}
\emailAdd{pawel.klimek@cern.ch}

\abstract{ This article reports the laser calibration of the hadronic Tile Calorimeter of the ATLAS experiment in the LHC Run~2 data campaign. The upgraded Laser~II calibration system is described. The system was commissioned during the first LHC Long Shutdown, exhibiting a stability better than 0.8\% for the laser light monitoring. The methods employed to derive the detector calibration factors with data from the laser calibration runs are also detailed. These allowed to correct for the response fluctuations of the 9852 photomultiplier tubes of the Tile Calorimeter with a total uncertainty of 0.5\% plus a luminosity-dependent sub-dominant term. Finally, we report the regular monitoring and performance studies using laser events in both standalone runs and during proton collisions. These studies include channel timing and quality inspection, and photomultiplier linearity and response dependence on anode current.}


\begin{document}
\maketitle
\flushbottom

\section{Introduction}
\label{sec:intro}
The ATLAS Tile Calorimeter (TileCal)~\cite{TCAL-2010-01} is the central hadronic calorimeter of the ATLAS experiment~\cite{PERF-2007-01} at CERN's Large Hadron Collider (LHC). The TileCal is a scintillator-based calorimeter employing photomultiplier tubes (PMTs) to measure the scintillation light. The TileCal is crucial to identify and measure the energy and direction of hadronic jets, provides information for the online trigger system and participates in the reconstruction of the missing transverse momentum associated to weakly-interacting particles. Thus, the TileCal plays a central role in the reconstruction of collision events for subsequent physics analyses.

The stability and resolution of the calorimeter response are parameters with direct impact on the precision of the reconstruction of jets and missing energy by the ATLAS experiment. In particular, the design characteristics of the experiment required a jet resolution of $\Delta E/E=50\%/\sqrt{E} + 3\%$ in the central region~\cite{ATLAS-TDR-03}. To achieve these parameters, TileCal is equipped with dedicated systems that allow to monitor the different components of the detector and calibrate its energy measurements. These procedures were conducted during the LHC Run~1 and Run~2 data taking campaign, contributing to a good operation and performance of the TileCal~\cite{TCAL-2017-01}. An important aspect was to correct for the response variation of the PMTs, achieved with a laser system. This article describes the laser calibration of the calorimeter in the LHC Run~2 data taking campaign. A previous report about the laser calibration in Run~1 can be found in Ref.~\cite{bib:laser_run_1}.

The calorimeter is briefly described in Section~\ref{sec:tilecal} and the Laser~II system operating during Run~2 is detailed in Section~\ref{sec:laser}. A major upgrade of the system employed in Run~1 was performed in 2014 and this is reported here. The description of the calibration procedure is presented in Section~\ref{sec:calibration}. Section~\ref{sec:laseringap} describes the monitoring of the calorimeter timing and PMT dependence on anode current with laser pulses fired during physics runs. Channel quality monitoring and studies of PMT linearity based on laser calibration data are reported in Section~\ref{sec:monitoring}. Finally, conclusions are drawn in Section~\ref{sec:conclusion}.


\section{The ATLAS Tile Calorimeter}
\label{sec:tilecal}
\newcommand{\AtlasCoordFootnote}{
ATLAS uses a right-handed coordinate system with its origin at the nominal interaction point (IP)
in the centre of the detector and the \(z\)-axis along the beam pipe.
The \(x\)-axis points from the IP to the centre of the LHC ring,
and the \(y\)-axis points upwards.
Cylindrical coordinates \((r,\phi)\) are used in the transverse plane, 
\(\phi\) being the azimuthal angle around the \(z\)-axis.
The pseudorapidity is defined in terms of the polar angle \(\theta\) as \(\eta = -\ln \tan(\theta/2)\).
Angular distance is measured in units of \(\Delta R \equiv \sqrt{(\Delta\eta)^{2} + (\Delta\phi)^{2}}\).}

\subsection{Detector overview}

\begin{figure}[htbp]
\begin{center}
    \includegraphics[width=0.5\textwidth]{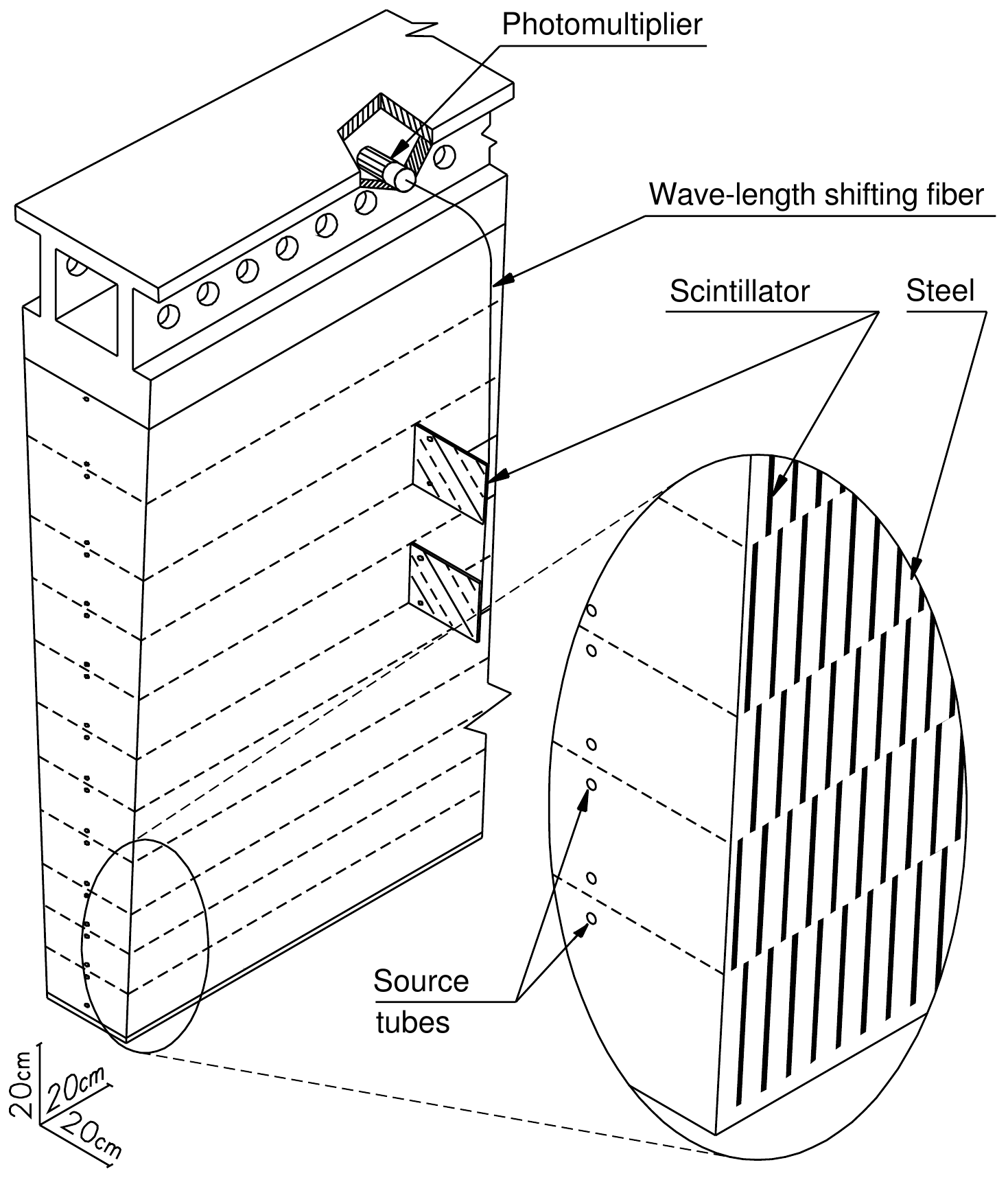}
    \caption{Sketch of a TileCal module, showing the scintillation light readout from the tiles by wavelength-shifting optical fibres and photomultiplier tubes (PMTs).}\label{fig:TileCalReadout}
\end{center}
\end{figure}

The TileCal is a non-compensating sampling calorimeter that employs steel as absorber material and scintillating tiles constituting the active medium placed perpendicular to the beam axis. The scintillation light produced by the ionising particles crossing the detector is collected from each tile edge by a wavelength-shifting (WLS) optical fibre and guided to a photomultiplier tube, see Figure~\ref{fig:TileCalReadout}. 

\begin{figure}[htbp]
\begin{center}
    \includegraphics[width=1.\textwidth]{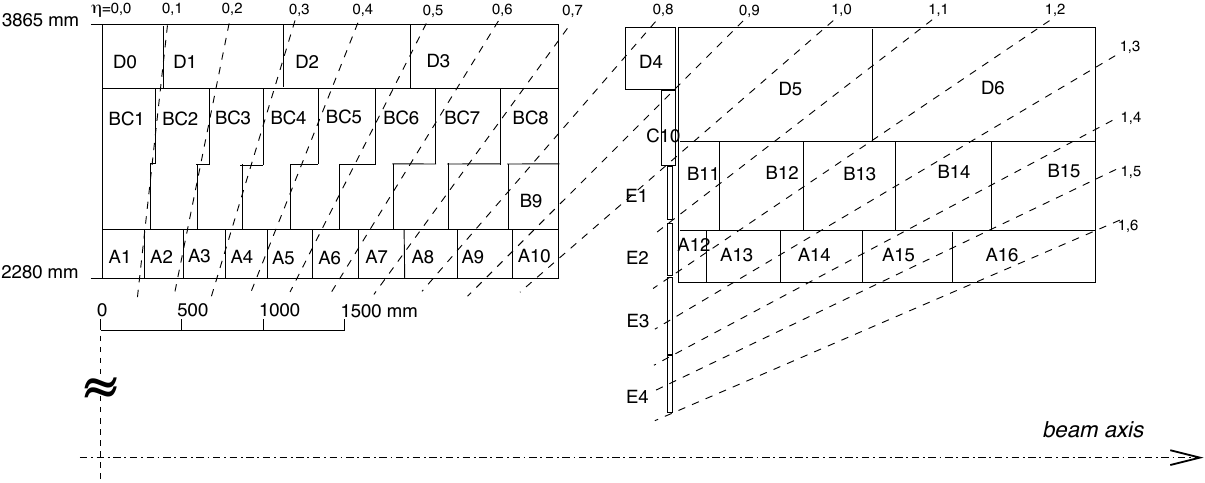}
    \caption{Scheme of the TileCal cell layout in the plane parallel to the beam axis, on the positive $\eta$ side of the detector. The single scintillators E1 and E2 (gap cells), and E3 and E4 (crack cells) located between the barrel and the end-cap are also displayed.}
\label{fig:TileCalSegmentation}
\end{center}
\end{figure}

The calorimeter covers a pseudorapidity range of $|\eta| < 1.7$ and is divided into three segments along the beam axis: one central long barrel (LB) section that is 5.8~m in length ($|\eta| < 1.0$), and two extended barrel (EB) sections ($0.8 < |\eta| < 1.7$) on either side of the LB that are each 2.6~m long\footnote{\AtlasCoordFootnote}. Full azimuthal coverage around the beam axis is achieved with 64 wedge-shaped modules, each covering $\Delta \phi = 0.1$ radians. 
Moreover, these are radially separated into 3 layers: A, B/BC and D. The readout cell units at each module are defined by the common readout of bundles of WLS fibres through a single PMT (see Figure~\ref{fig:TileCalReadout}). The cell mapping in the $(r,\eta)$-plane is sketched in Figure~\ref{fig:TileCalSegmentation}. The great majority of the cells have an independent readout by left/right PMTs for each cell side providing redundancy for the cell energy measurement. Additionally, single scintillator plates are placed in the gap region between the barrels (E1 and E2 cells) and in the crack in front of the ATLAS electromagnetic calorimeter End-Cap (E3 and E4 cells).

The data acquisition system of the TileCal is split into four partitions, the ATLAS A-side ($\eta > 0$) and C-side ($\eta < 0$) for both the LB and EB, yielding four logical partitions: LBA, LBC, EBA, and EBC. In total, the TileCal has 5182 cells and 9852 PMTs. 

The PMT model Hamamatsu R7877 is used, which is a special customised 8-stage fine-mesh version of Hamamatsu R5900 with nominal gain of $10^5$ for a high voltage of $\sim$650~V and a dark current of about 0.3 nA~\cite{ATLAS-TDR-03,Crouau:1997tka}. The photocathode material is Bialkali with a borosilicate window, and the quantum efficiency is close to 18\% at 480~nm. The typical anode rise-time corresponds to 1.4 ns, whereas the typical transit time has a spread of 0.3 ns (for a transit time of several ns).

The front-end electronics~\cite{Anderson:2005ym} receive the electrical signals from the PMTs, which are shaped, amplified with two different gains in a 1:64 proportion, and then digitised at 40~MHz sampling frequency~\cite{Berglund:2008zz}.
The bi-gain system is used in order to achieve a 16-bit dynamic range using 10-bit ADCs.
The digital samples are stored in a pipeline memory. Upon ATLAS Level~1~\cite{TRIG-2019-04} trigger decision, seven signal samples are sent to the detector back-end electronics for the reconstruction of the signal amplitude. Complementarily, the PMT signals are integrated over a long period of time (10--20~ms) with analog integrator electronics to measure the energy deposited during caesium calibration scans and the charge induced by proton--proton ($pp$) collisions.

\subsection{Signal reconstruction}
\label{sec:signal_energy_reco}

In each TileCal channel, an analog electrical signal is sampled with seven measurements at 25~ns spacing synchronised with the LHC master clock. These samples are referred to as $S_i$, where $1\leq i \leq 7$, and are in units of ADC counts. Depending on the amplitude of the pulse, either High or Low Gain is used to maximise the signal to noise ratio while avoiding saturation. To reconstruct the sampled signal produced during physics runs, the Optimal Filtering (OF) method is used in the Tile Calorimeter~\cite{Cleland:2002rya, Fullana:816152}. The method linearly combines the samples $S_i$ to calculate the amplitude $A$, phase $\tau$ with respect to the 40 MHz clock and pedestal $p$ of the pulse:

\begin{equation}
A=\sum_{i=1}^{n=7}a_iS_i,\hspace{3em}
A\tau=\sum_{i=1}^{n=7}b_iS_i,\hspace{3em}
p=\sum_{i=1}^{n=7}c_iS_i
\label{eq:of}
\end{equation}

where $a_i$, $b_i$ and $c_i$ are linear coefficients optimised to minimise the bias on the reconstructed quantities introduced by the electronic noise. The normalised pulse shape function, taken as the average pulse shape from test beam data, is used to determine the coefficients. Separate functions are defined for high and low gain. The pulse shape and coefficients are stored in a dedicated database for calibration constants.

The system clock in each digitiser~\cite{Berglund:2008zz} is tuned so that the signal pulses, originating from collisions at the interaction point, 
peak at the central (fourth) sample, synchronous with the LHC clock. The reconstructed value of $\tau$ represents the small time phase in ns between the expected pulse peak and the time of the actual reconstructed signal peak, arising from fluctuations in particle travel time and uncertainties in the electronics readout.

To reconstruct the signals produced in each TileCal channel by the laser calibration system, the same OF method is used as during the physics runs. 
In this case, the pulse shape function corresponding to the signal produced by laser is used to calculate the linear coefficients $a_i$, $b_i$ and $c_i$ from Equation~\eqref{eq:of}. 

%
%
%
%

\subsection{Energy reconstruction and calibration}

\begin{figure}[htbp]
\centering
\includegraphics[width=1.\textwidth]{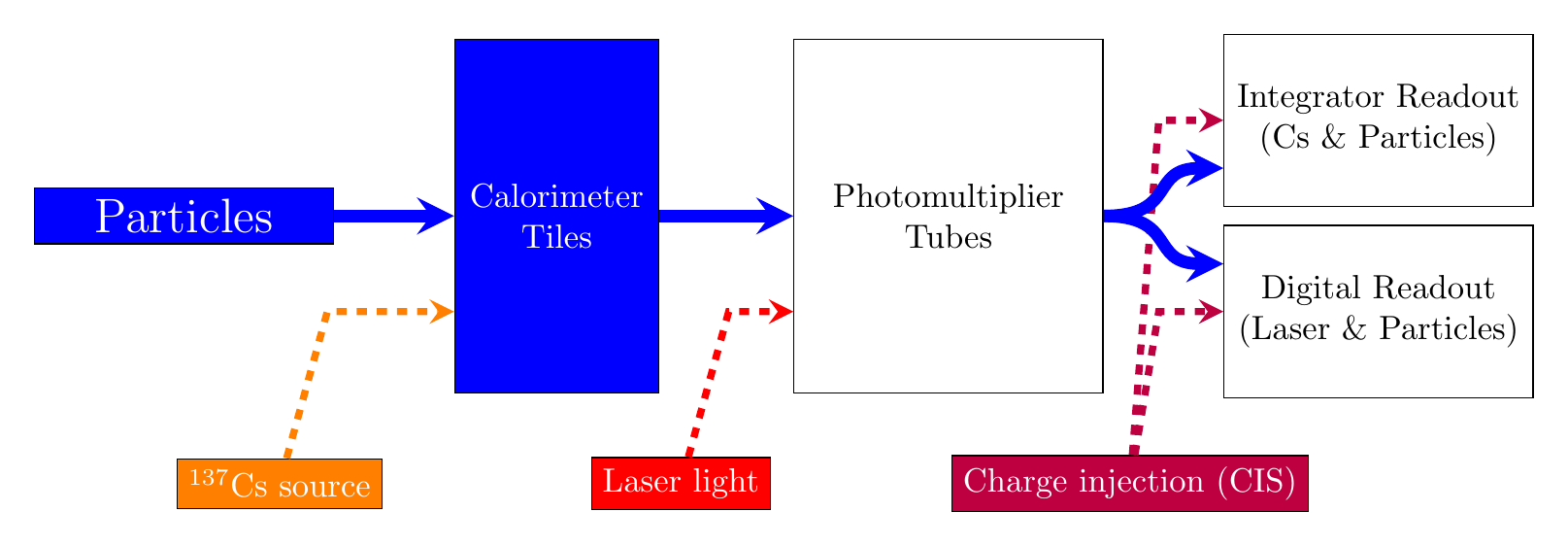}
\caption{The signal paths for each of the three calibration systems used by the TileCal. The physics signal is denoted by the thick solid line and the path taken by each of the calibration systems is shown with dashed lines.}
\label{fig:TileCalCalibrationChain}
\end{figure}

At each level of the TileCal signal reconstruction, there is a dedicated calibration system to monitor the behaviour of the different detector components. 
Three calibration systems are used to maintain a time-independent electromagnetic (EM) energy scale, and account for variations in the hardware and electronics. A movable caesium radioactive $\gamma$-source calibrates the optical components and the PMTs but not the front-end electronics~\cite{Blanchot:2020lyh}. The laser system monitors the PMTs and front-end electronic components used for collision data. The charge injection system (CIS) calibrates the front-end electronics~\cite{TCAL-2010-01}. Figure~\ref{fig:TileCalCalibrationChain} shows a flow diagram summarising the different calibration systems along with the paths followed by the signals from different sources. These three complementary calibration systems also aid in identifying the source of problematic channels. Moreover, the minimum-bias currents (''Particles`` in Figure~\ref{fig:TileCalCalibrationChain}) are used to validate response changes observed by the caesium calibration system. 

In each TileCal channel, the signal amplitude $A$ is reconstructed in units of ADC counts using the OF algorithm defined in Equation~\eqref{eq:of}. The reconstructed energy $E$ in units of GeV is derived from the signal amplitude as follows: 

\begin{equation}
E\ [\mathrm{GeV}]=\frac{A\ [\mathrm{ADC}]}{f_{\mathrm{pC}\to \mathrm{GeV}}\cdot f_{\mathrm{Cs}}\cdot f_{\mathrm{Las}} \cdot f_{\mathrm{ADC}\to \mathrm{pC}}}
\label{eq:channelEnergy}
\end{equation}

where each $f_i$ represents a calibration constant or correction factor. 
The factors can evolve in time because of variations in PMT high voltage, stress induced on the PMTs by high light flux, PMT ageing or radiation damage to scintillators. The calibration systems are used to monitor the stability of these factors and provide corrections for each channel.

The $f_{\mathrm{pC}\to \mathrm{GeV}}$ conversion factor is the absolute EM energy scale constant measured in test beam campaigns~\cite{testBeam}. $f_{\mathrm{ADC}\to \mathrm{pC}}$ is the charge to ADC counts conversion factor determined regularly by charge injection, and the remaining factors, $f_{\mathrm {Cs}}$ and $f_{\mathrm{Las}}$, are calibration factors measured frequently with the TileCal calibration systems (f.i., laser calibration runs are taken on a daily basis). These are updated in the database according to an \textit{interval of validity} (IOV) and used by the data preparation software to keep the cell energy response stable over time. The IOV has a start and end run identifier, between which the stored conditions are valid and applicable to data, and is also stored in the database.

The calibration activities and the precision on the calibration factors has direct impact on the resolution of the cell energy measurement which reflects on the precision of the reconstruction of the ATLAS high-level physics objects used by physics analysis. In Run 2, the TileCal calibration contributed attain a measured jet energy resolution of 3.5\% for central jets of very high transverse momenta~\cite{ATLAS:2020cli}, matching the design goals of the experiment. In particular, the precision of the laser calibration is discussed in Section~\ref{sec:uncertainty}.


\section{The Laser II calibration system}
\label{sec:laser}
\subsection{Laser calibration system}

The TileCal laser system is installed in the ATLAS main service cavern, USA15, located about 100~m away from the detector. In summary, it consists of a laser source, light guides and beam expanders, an optical filter wheel to adjust the light intensity, and beam splitters to dispatch the light to the Tile Calorimeter PMTs through 400 clear optical fibres, 100 to 120~m long. In addition, the system is equipped with a calibration setup designed to monitor the light in various points of the dispatching chain, and with dedicated control and acquisition electronics boards.

The original Laser~I system used during the LHC Run~1 operation~\cite{bib:laser_run_1} was upgraded during the LHC Long Shutdown~1 to a newer version, referred to as Laser~II~\cite{Gris:2016uqb,vanWoerden:2016luu}, which is used since the beginning of the Run~2 data taking. The main purpose of the laser upgrade was to overcome the shortcomings observed in the previous system and improving the precision and stability marks. The essential aspects to improve were the stability of the beam expander distributing light to the 400 clear fibres, the photodiodes grounding, reproducibility of the filter wheel position, the electronics, and the overall estimate of the laser light injected into the PMTs of the calorimeter.

\subsection{Upgraded Laser~II system}

The Laser~II system can be described by six main functional blocks: the optics box, the optical filter patch panel, the photodiode box, the PHOCAL (PHOtodiode CALibration) module, a PMT box and a VME crate featuring the LASCAR (LASer CAlibration Rod) electronics card. These blocks are briefly described below, highlighting the upgrades with respect to the previous Laser~I version.

\begin{figure}[htbp]
\begin{center}
    \includegraphics[width=0.8\textwidth]{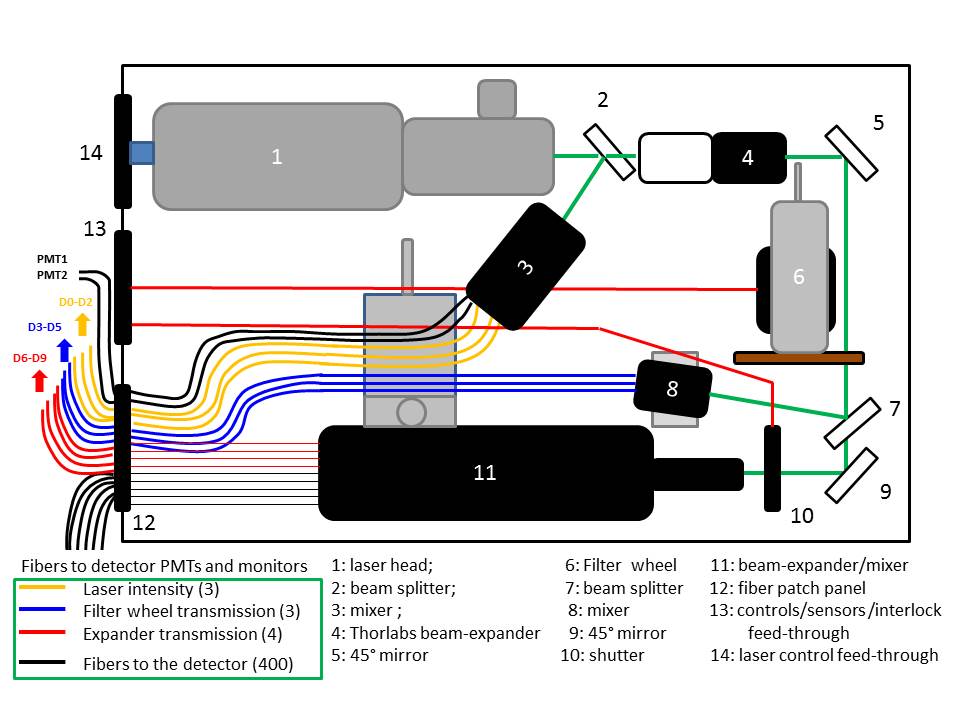}
    \caption{Scheme of the Laser~II optics box, depicting the internal elements and optical paths.}\label{fig:optics_box}
\end{center}
\end{figure}

\begin{figure}[htbp]
\begin{center}
     \includegraphics[width=0.45\textwidth]{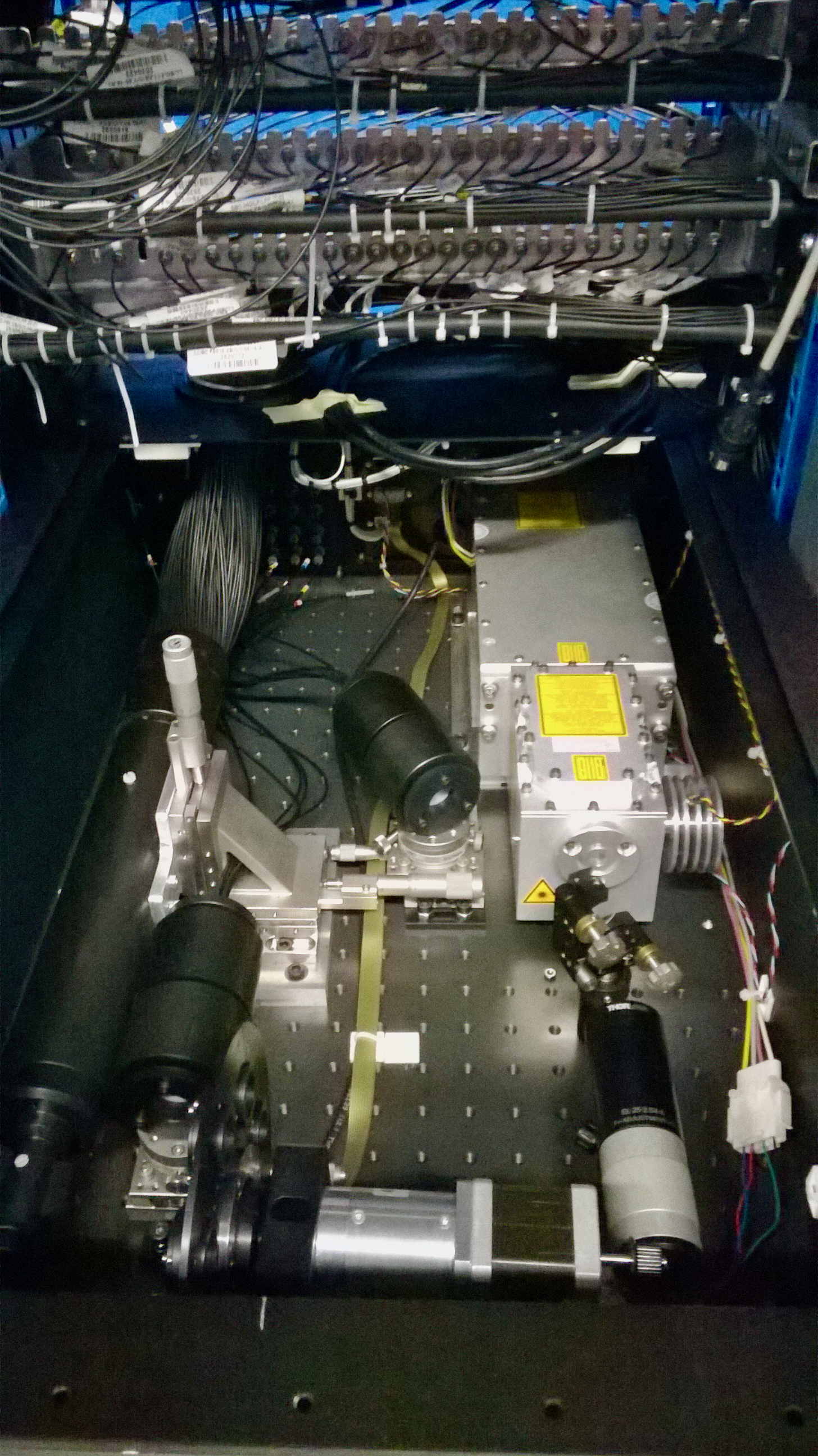}
    \caption{Picture of the optics box (cover removed) placed on the anti-vibration rails and coupled to the fiber bundle.}\label{pic:optics_box}
\end{center}
\end{figure}

\subsubsection*{Optics box}

The light source of the laser system, installed in the so-called optics box, is a commercial Q-switched diode-pumped solid state laser manufactured by SPECTRA-PHYSICS~\cite{ref:laser}, kept from the predecessor system. The frequency doubler permits the infrared laser to emit 532~nm green light, close to the wavelength of the light coming from the detector WLS fibres, peaked at 480~nm. The time width of the individual pulses generated by the laser is 10~ns. Besides the laser source, the optics box houses also the main optical components acting on the laser beam across the light path, as depicted in Figure~\ref{fig:optics_box}. A picture of the laser box is shown in Figure~\ref{pic:optics_box}.

A beam splitter is located at the output of the laser cavity. It divides the laser primary beam into two parts: a small fraction of the light is sent back to a light mixer and the major part is transmitted through a beam expander and a 45\textdegree{} dielectric mirror to a filter wheel. The light exiting the mixer is collected by five clear optical fibres: two are coupled to the PMT box to the two PMTs responsible for generating the trigger signal for the Laser~II data acquisition (DAQ); and three are connected to the monitoring photodiodes (D0 to D2) located in the photodiode box. 

In the expander, the beam spot is expanded from 700~$\mu$m to 2~mm, reducing the light power density to the forthcoming optical elements. The light reflected by the following mirror passes through a motorised filter wheel hosting eight neutral density filters with varied optical densities, with the filter transmissions ranging between 100\% (no filter) and 0.3\%. The combination of this transmission variation and the range of intensities where the laser operation is stable allows to calibrate the TileCal PMTs in an equivalent cell energy range of 500~MeV to 1~TeV. 

The light transmitted by the selected filter is fed into a light mixer by a beam splitter placed downstream the wheel. Three clear fibres routed to the photodiode box collect the light for monitoring (diodes D3 to D5). A second 45\textdegree{} dielectric mirror reflects the light through a shutter into the final beam expander where the laser light is finally dispatched to the detector by 400 clear fibres. Four fibres route the light output of the expander to the photodiode box to monitor its transmission (photodiodes D6 to D9). 

The 400 long clear fibres are bundled together and transfer the light coming out of the optics box to the TileCal modules (one fibre for two half modules of the central Long barrel, one fibre for each half module per extended barrel and 16 spare fibres). The association between TileCal PMTs and the clear fibre in the bundle is as follows:
\begin{itemize}
\item Long Barrel: one fibre per full LB module, for even PMT numbers in A side and odd numbers C side. Conversely, another fibre for the same LB module, for odd PMT numbers in A side and even numbers in C side.
\item Extended Barrel: one fibre per module per EB side for the even number PMTs and another one for the odd PMTs.
\end{itemize}
Inside each detector module an optical system composed of a light mixer in air dispatches the light to each PMT with individual clear fibres.

This optics box comprises major upgrades with respect to the previous system. Now, a compact design of the optical layout fully includes all the optical elements in one single box, whereas in the Run~1 system the optical elements were located into two different boxes optically connected with a liquid fiber. The optics box is set in an horizontal position to minimise the dust accretion on optical parts and to ease interventions and is mounted on an anti-vibration system, improving beam stability. The final beam expander is new. It was re-designed to improve uniformity in the distribution of the 2~mm beam spot across the 400 fibres' bundle, which has a circular surface of 30~mm diameter. Finally, the system now permits a better estimate of the laser light injected in the calorimeter through a redundant monitoring of the light transmitted in different points of the optical line with 10 photodiodes.

\subsubsection*{Optical filters}

A patch panel with ten optical filters is used to adjust the intensity of the light read by each of the monitoring photodiodes in the photodiode box. In this set up, each one of the ten optical fibres reading out the light at the various points of the beam path in the optics box (after the laser head, after the filter wheel, and at the output of the beam expander) is coupled to a given optical filter in the patch panel. The optical density of the filters range from 0.5 to 2.5 and are such that for each light point probed there is always at least two filters of equal density, providing a redundant light intensity probe for the monitoring photodiodes.

\subsubsection*{Photodiode box}

The photodiode box is a rack containing a set of ten modules, each composed of a Si PIN photodiode (Hamamatsu S3590-08~\cite{ref:photodiode}) coupled to a pre-amplifier, a control card, and a charge injection card to inject an electrical charge into the ten pre-amplifiers. A set of two fibres is connected to the rear end of the photodiode box, in front of each photodiode. One fibre conveys the laser light for monitoring and is connected from the patch panel with the optical filters. The other one comes from the PHOCAL module, where LED light is injected to assess the stability of the photodiodes. In order to minimise the photodiodes' response dependency on the temperature, the temperature in rack is controlled by the water and fan cooling system. The temperature of each photodiode is monitored and kept constant at approximately 30~$^\circ$C with a long term stability below 1~$^\circ$C.

\subsubsection*{PHOCAL module}

This module implements a redundant internal calibration scheme using a blue LED (Nichia blue NSPB520S $\lambda$=470~nm~\cite{ref:led}) to monitor the ten photodiodes of the Laser~II system. The calibration light is simultaneously transmitted to a reference photodiode (Hamamatsu S2744, active area: 10x20 mm2, spectral response range from 320 to 1100~nm~\cite{ref:bigphoto}) providing the signal for the photodiodes' response normalisation. PHOCAL also contains a radioactive source of $^{241}$Am, releasing mostly $\alpha$ particles of 5.6 MeV with an activity of 3.7~kBq. This source ensures the monitoring of the reference photodiode. This module is an addition with respect to the Laser~I system installed in Run~1, where the existing photodiodes were all monitored with a moveable scheme of the $^{241}$Am source.

\subsubsection*{PMT box}

The PMT box contains two PMTs (Hamamatsu R5900~\cite{ref:pmt}) reading out two optical fibres from the optical box. These provide the trigger signal for the Laser~II acquisition system when the laser is flashing. The PMT box also includes a control module used to drive the shutter and the filter wheel in the optics box.

\subsubsection*{LASCAR electronics}

\begin{figure}
\centering
\includegraphics[height=7cm]{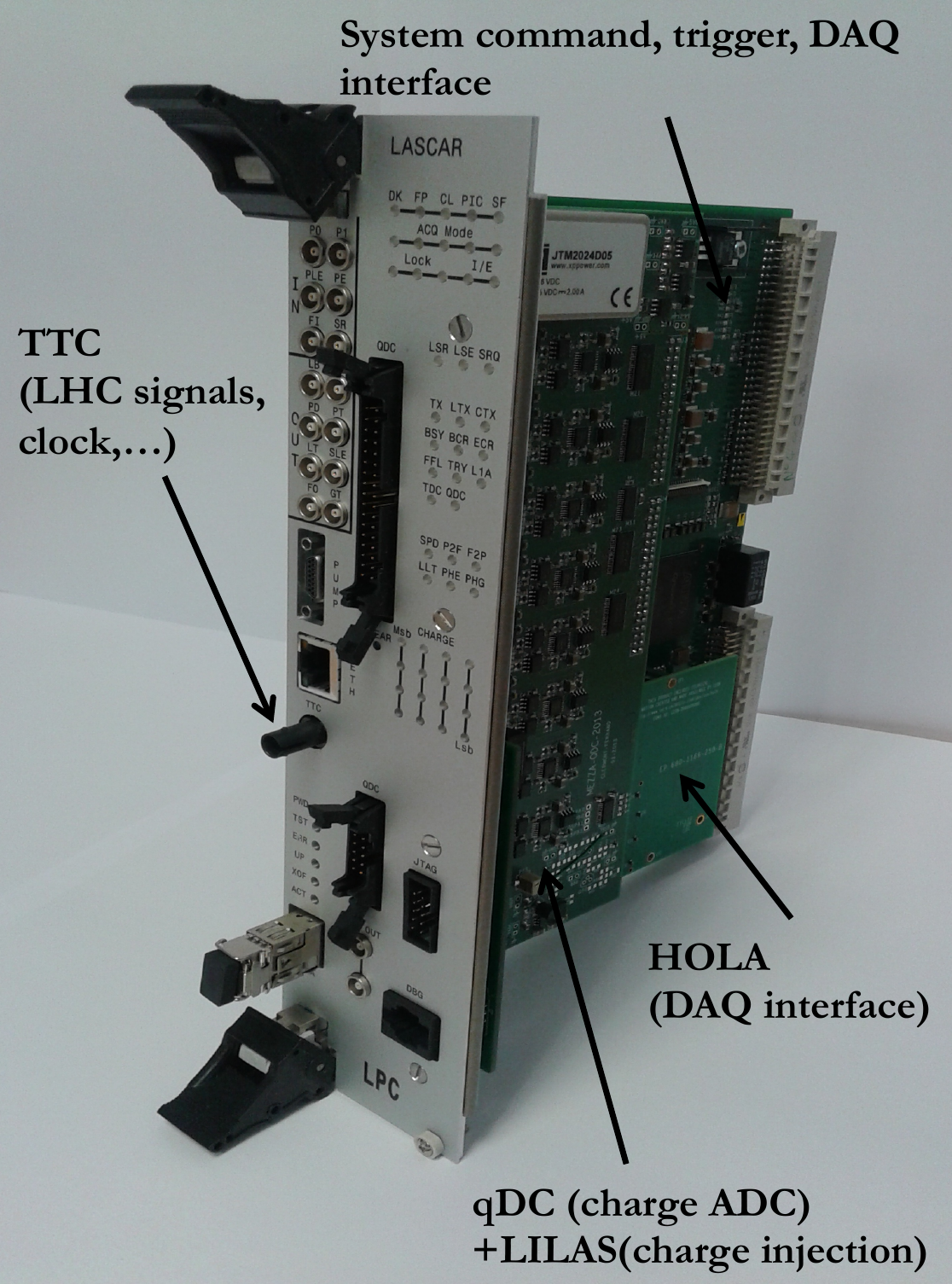}\quad
\includegraphics[height=7cm]{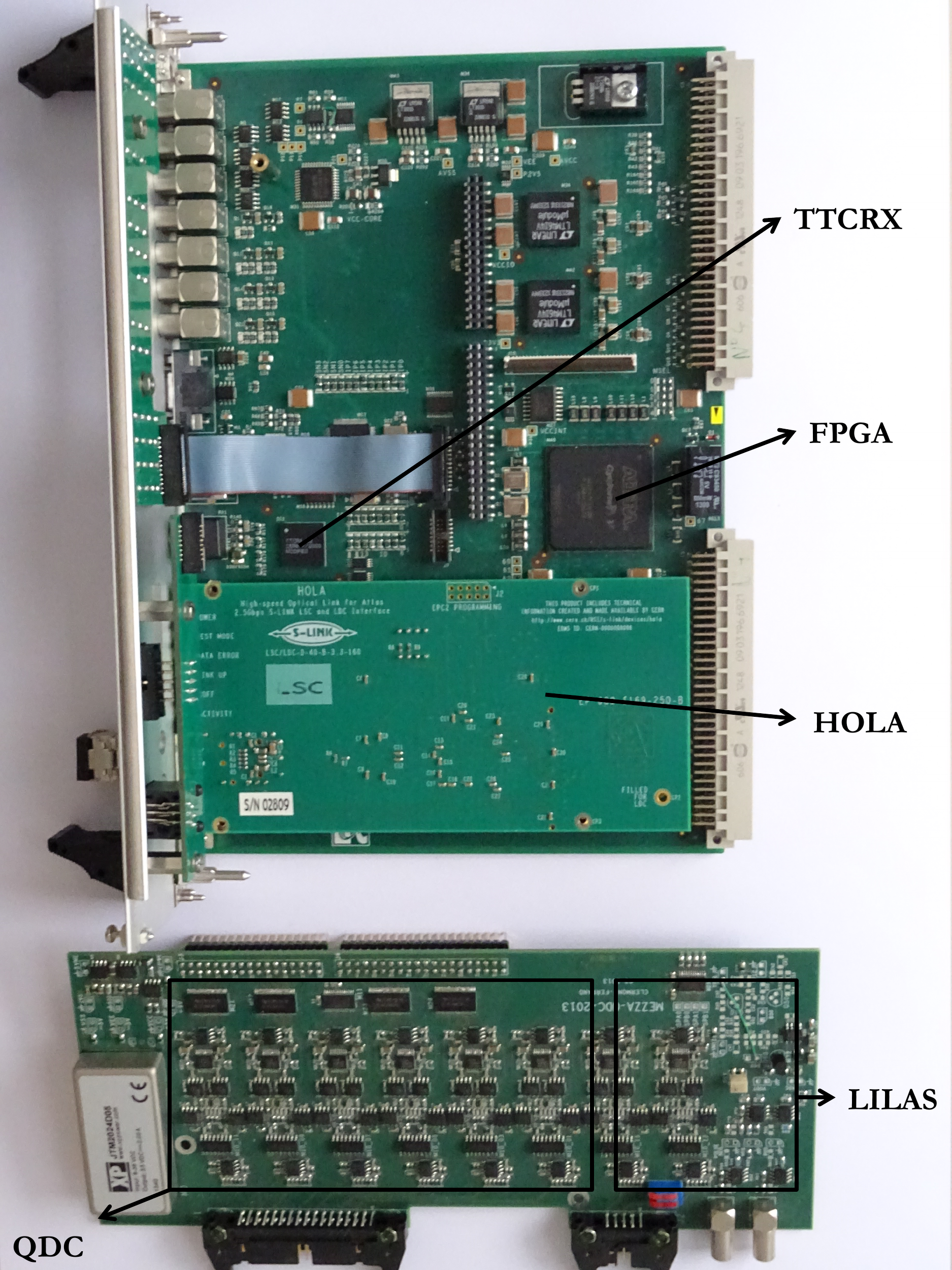}
\caption{Views of the LASCAR card.}\label{fig:laslascar}
\end{figure}

LASCAR, viewed in Figure~\ref{fig:laslascar}, is the electronics board for the acquisition and control of the Laser~II system~\cite{pgris_laserii}. It digitises the analog signals (from the eleven photodiodes, the two PMTs and the charge injection system), contains a chip to retrieve the LHC clock signal, makes the interface to drive the laser and contains the module for charge injection and monitoring of the pre-amplifier and digitisation chain of the photodiodes.

The LASCAR board is housed in a VME crate. It's central brain, a Field Programmable Gate Array (FPGA) Cyclone V manufactured by ALTERA~\cite{ref:altera-cyclone}, with 150K logic elements and 7.5~MB of RAM, controls and provides the interface to the main components:

\begin{itemize}

\item \textbf{Charge ADC (QDC):} 32 channel 14-bit\footnote{One of the bits is used to indicate whether the signal is positive or negative, resulting in an effective 13-bit dynamic range for a maximum integrated charge of 2000~pC.} QDC that performs a 500~ns integration and digitization of the analog input charge signals coming from the eleven photodiodes, the two PMTs and from the charge injection system. Prior to the QDC, the analog signal pass through a charge amplifier circuit with two possible gains ($\times 1$ and $\times 4$).
\item \textbf{LILAS (LInearity LASer) card:} This module is responsible for injecting a known charge into the readout electronics (photodiodes pre-amplifiers and digitisation) to monitor its linearity and stability with time. A digital signal from the FPGA is converted to an electric charge by the LILAS 16-bit DAC. The charge is then injected directly into a QDC channel and distributed to the PHOCAL and Photodiode box through lemo cables.
\item \textbf{Time-to-Digital Converter (TDC)}: A TDC is used to measure the Laser time response as a function of its intensity. The device has two channels and a time resolution of 280~ps. LASCAR is equipped with a delay system to insure the adequate laser pulse timing irrespective of laser amplitude.
\item \textbf{Timing, Trigger and Control Receiver (TTCrx)}: The TTCrx is an ASIC chip that receives the LHC signals relative to bunch crossing, event counter reset and trigger.
\item \textbf{HOLA}: The High-speed Optical link for ATLAS was conceived to send data fragments via optical fibre to the Read Out System of ATLAS upon receiving a Level 1 Accept trigger from the ATLAS central DAQ.
\item \textbf{LASER Interface:} This mixed analog and digital board is used to control the laser head. The laser intensity is set through an analog signal (0 to 4~V) and the trigger is set with a TTL signal.
\end{itemize}

\subsection{Operating modes}

The Laser~II system can be operated independently as a stand-alone system or integrated in the ATLAS detector data acquisition framework.

\subsubsection*{Stand-alone operating mode}

The stand-alone operation of the Laser~II allows to verify that the system is responding as expected, to monitor its stability and to perform its internal calibration. In this internal calibration mode, LASCAR controls the Laser~II components, switching the shutter off by default, without sending any laser pulse to TileCal PMTs. The following running modes are possible:

\begin{itemize}
\item \textbf{Pedestal mode:} This mode is used to measure a high number of events when no input signal is injected (from the laser, the LED or the radioactive source).
\item \textbf{Alpha source mode:} This mode is used to measure the response of the reference photodiode in the PHOCAL module to the $\alpha$ particles emitted by the $^{241}$Am source.
\item \textbf{LED mode:} In this mode, the LED signal is transmitted to all the photodiodes, including the reference photodiode, via optical fibres. It allows to probe the stability of the photodiodes used to monitor the laser light.
\item \textbf{Linearity mode:} This mode resources the LILAS card to inject a known electrical charge into the preamplifiers of the photodiodes in order to assess the stability of the electronics. It also allows to vary the injected charge to evaluate the linearity of the readout electronics.
\item \textbf{Laser mode:} In this mode, the laser signals of adjustable intensity are sent into the system. The light can be transmitted to the TileCal PMTs, depending on the status of the shutter located inside the optics box. 
\end{itemize}

A standard internal calibration run combines all the above running modes, starting with the pedestal mode. Once enough pedestal events have been recorded, LASCAR is switched to the next calibration mode, starting from the alpha mode, then the LED mode and the linearity mode increasing the injected charge from 0 to 60000 DAC counts ($\sim$1.9~pC) by steps of 10000 ($\sim$0.3~pC). Finally, the internal calibration ends with the laser mode.
This internal calibration sequence is ran approximately daily, before each laser calibration run described next.

\subsubsection*{ATLAS DAQ operating mode}

The ATLAS DAQ mode is the main operating mode of Laser~II. Its role is to calibrate the TileCal PMTs with laser light. To do so, the Laser~II DAQ is integrated within the global ATLAS DAQ infrastructure, which handles the readout of the PMT signal induced by the laser and the Laser~II run control. This mode is used in two ways:

\begin{itemize}
\item \textbf{Laser mode:} This is the main mode to perform the dedicated calibration runs, when the TileCal is operated independently of the remaining ATLAS detector. 
Laser pulses are sent to the calorimeter by request of the SHAre Few Trigger board (SHAFT) to LASCAR. 
Amplitudes of the signals produced by the photodiodes and the PMTs of the Laser~II system are sent back to the ATLAS DAQ by LASCAR. At the start of run, the filter wheel position and the laser intensity are configured and the shutter opened. These runs are taken on a daily basis, either in technical stops or during proton bunch inter-fill stops in collision data-taking.
\item \textbf{Laser-in-gap mode:} Laser pulses are emitted in empty bunch-crossings during standard physics runs of the LHC. Empty bunch crossings are those with no proton bunch and are separated from any filled bunch by at least five bunch crossings to ensure signals from collision events are cleared from the detector. The TileCal is synchronised with the other ATLAS sub-detectors. The light is fired only in exclusive periods of the beam orbit, where no collisions can occur. The SHAFT board sends a request to LASCAR at a fixed time with respect to the beginning of the LHC orbit to synchronise the laser pulses with the pre-defined orbits and ensure no overlap between laser and physics events. Upon pulse emission, a laser calibration request is sent to the ATLAS central trigger processor by the SHAFT interface. This arrangement synchronises the pulse emission and the TileCal readout with ATLAS DAQ in physics runs.

\end{itemize}

\subsection{Stability of the laser system}

\begin{figure}[htbp]
\begin{center}
\subfloat[Pedestal\label{fig:pedestal}]{\includegraphics[height=0.28\textwidth]{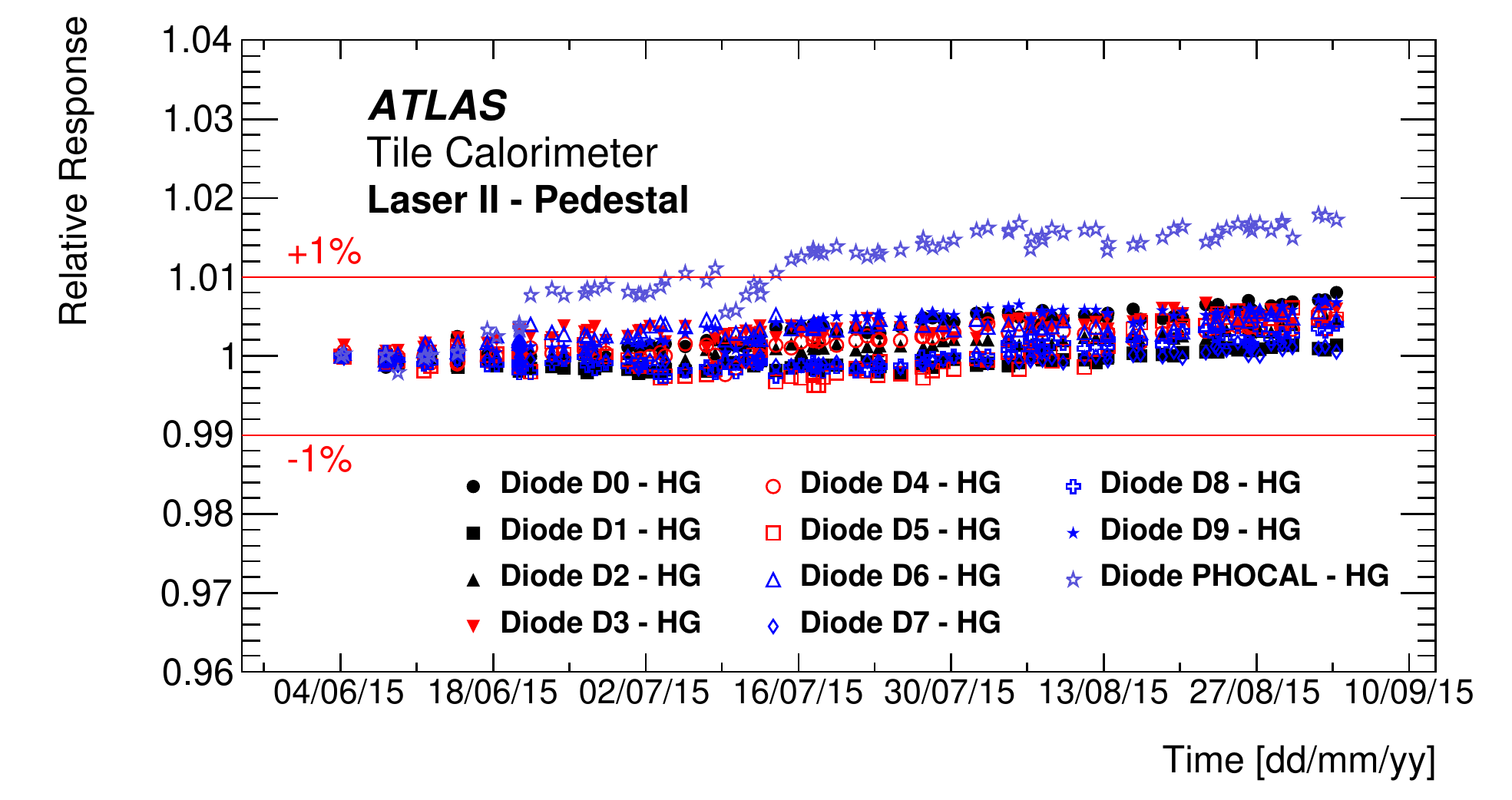}}
\subfloat[Charge Injection\label{fig:charge_injection}]{\includegraphics[height=0.28\textwidth]{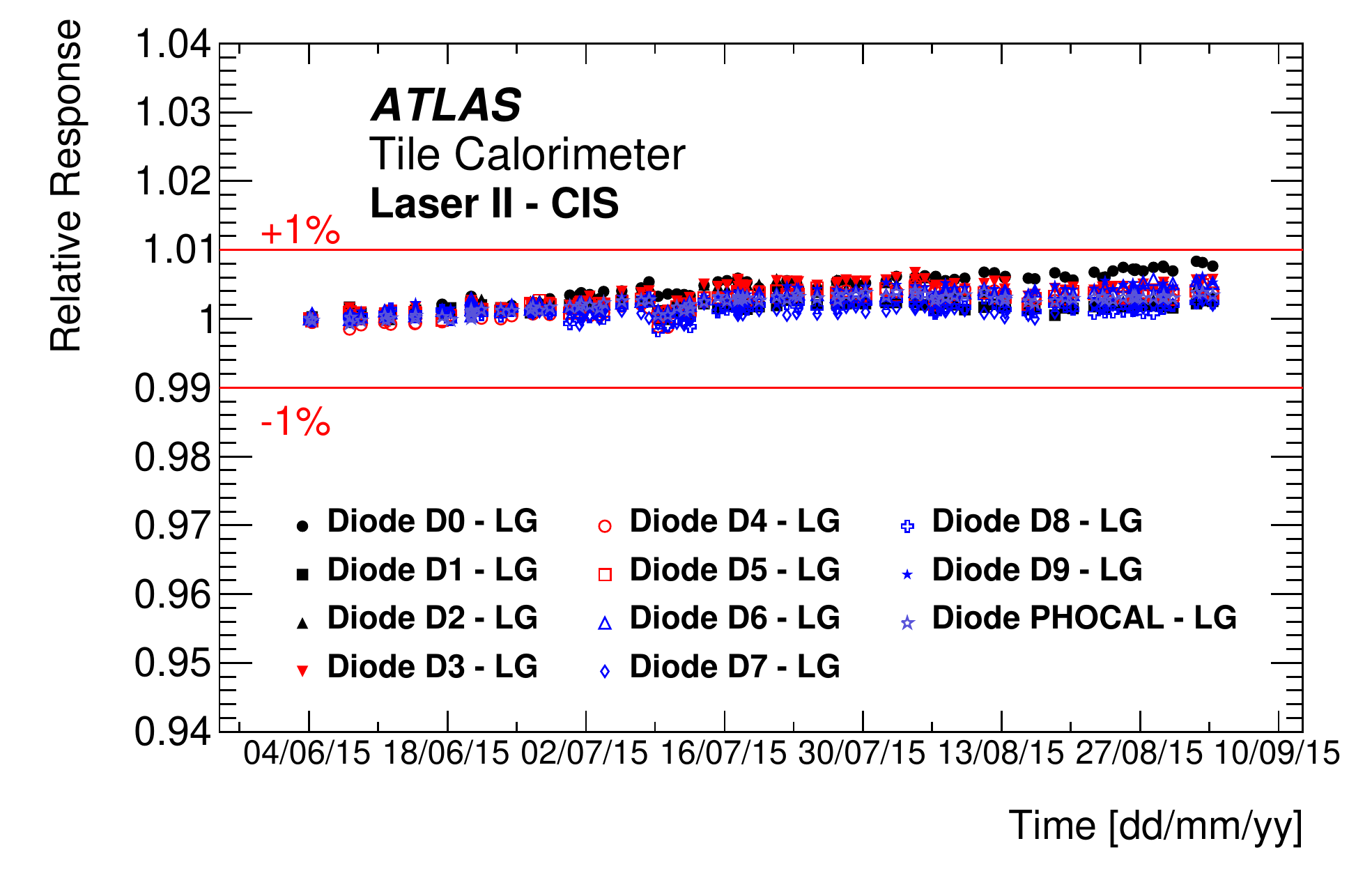}}\\
\subfloat[LED\label{fig:led}]{\includegraphics[height=0.28\textwidth]{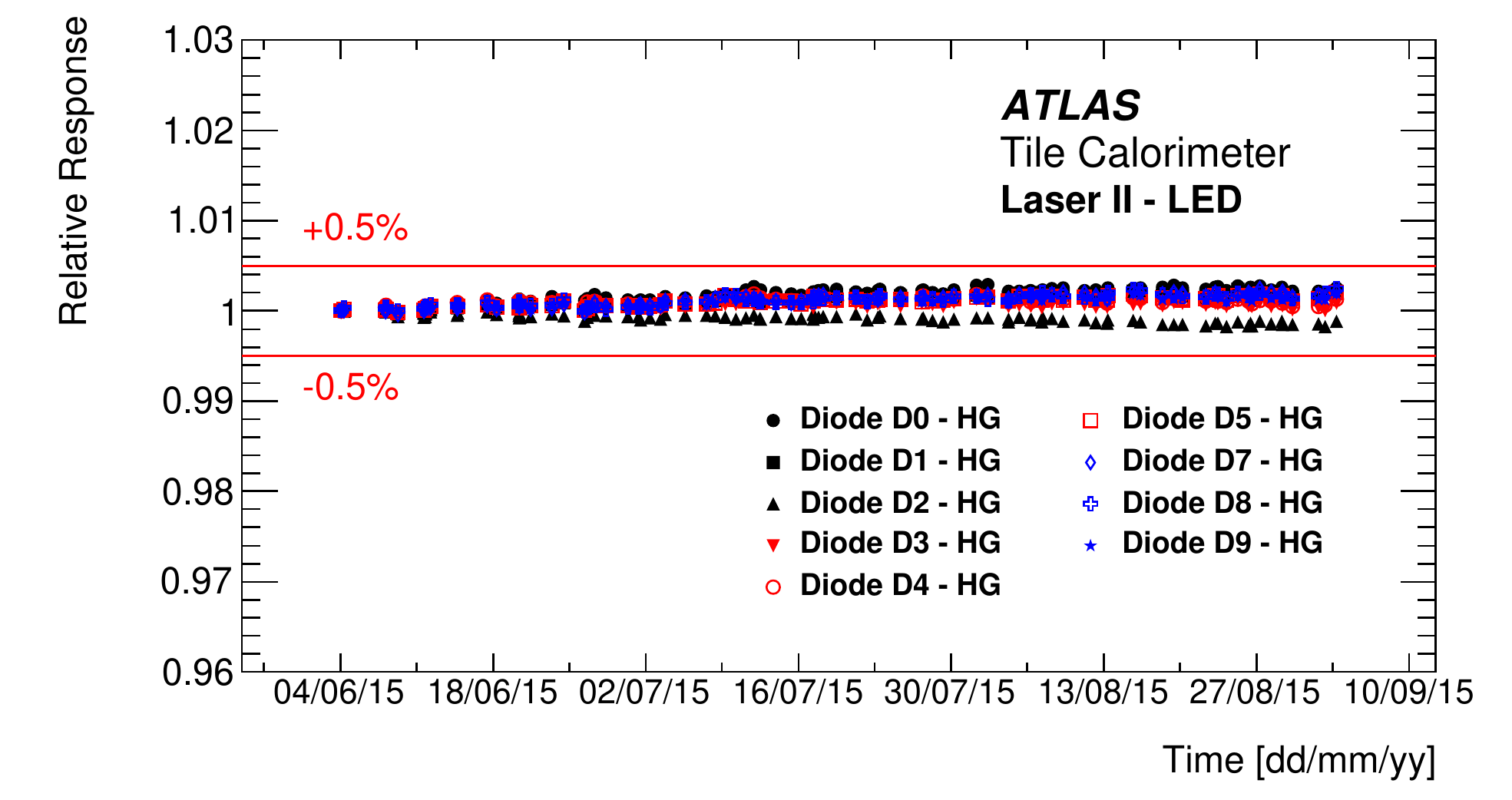}}
\subfloat[$^{241}$Am $\alpha-$source\label{fig:alpha}]{\includegraphics[height=0.28\textwidth]{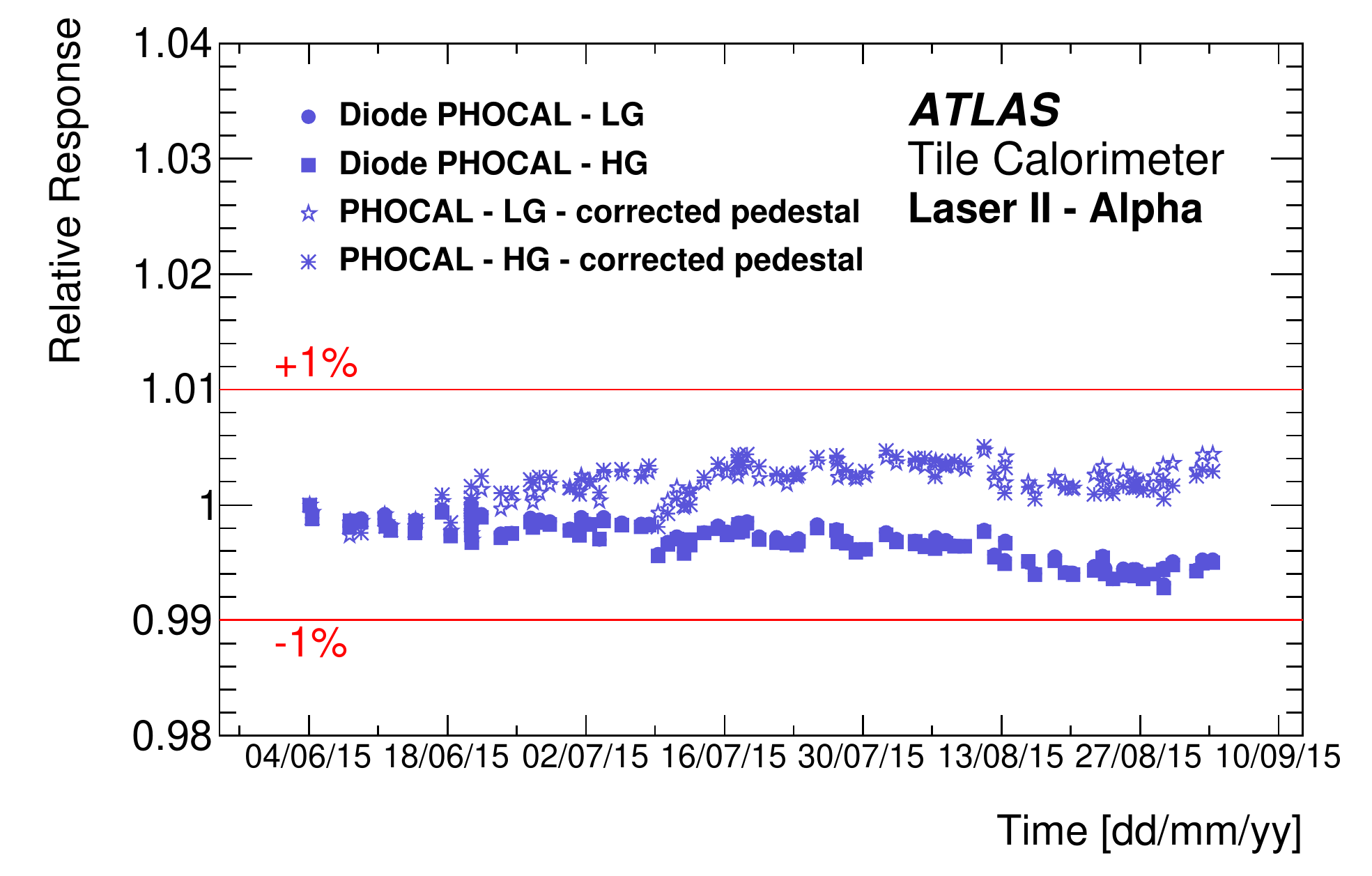}}
    \caption{Relative (a) pedestal and (b) charge injection signal for each photodiode (D0 to D9 and reference photodiode in PHOCAL) as a function of time in the course of the first three months of Run~2. The mean signal values are pedestal subtracted and normalised to the mean signal value of the first measurement. (c) Relative response to the PHOCAL LED for each photodiode (D0 to D9) as a function of time. Signal values are pedestal subtracted and normalised to the PHOCAL photodiode signal and to the mean signal value of the first measurement. (d) Relative response of the PHOCAL photodiode to the $^{241}$Am $\alpha$-source subtracting a constant pedestal or correcting the pedestal for the observed fluctuation in time. The mean signal values are normalised to the mean signal value of the first measurement and HG and LG refer to the High and Low Gain acquisition modes, respectively .}\label{fig:laser_stability}
\end{center}
\end{figure}

The installation of Laser~II involved a commissioning phase where the performance and the stability of the system were evaluated in the course of the first three months of Run~2~\cite{Scuri:2016ctn}. The main parameters to monitor are the ones obtained with the operation of the laser internal calibration mode. The measurements included the pedestal of the photodiodes, the response of the electronics to a known injected charge, the signal of the monitoring photodiodes in response to the PHOCAL LED pulses, and the response of the PHOCAL photodiode to the $^{241}$Am $\alpha$-source. The results are shown in Figure~\ref{fig:laser_stability}.

Figure~\ref{fig:pedestal} shows the average pedestal value recorded in the Laser II stand-alone acquisition mode in high gain for the monitoring photodiodes (D0 to D9) and for the PHOCAL photodiode. The data is normalised relatively to the first data point. The pedestals of the D0--D9 photodiodes are stable within 0.8\% during the commissioning period, whereas for the PHOCAL diode a maximum variation of 1.8\% is observed. 

The stability of the readout electronics response is obtained by injecting a constant charge of 171~pC, 256~pC or 342~pC in several runs across the considered time period. For each injected charge, the signal is acquired in low gain and in high gain. Figure~\ref{fig:charge_injection} shows the results obtained for a 256~pC injected charge in low gain readout. The data, normalised to the first day of data taking, are shown for the electronics channels corresponding to each photodiode and the pedestals are subtracted. All channels exhibit a consistent up-drift reaching 0.8\% at the most in the end of the data taking period. 

In Figure~\ref{fig:led}, the outcome of the photodiode monitoring with the PHOCAL LED in stand-alone high gain runs of Laser~II is presented. The values are normalised to the first data point and the pedestals are subtracted. The response of the photodiodes to the calibration light is very stable in time. The maximum fluctuations do not overcome 0.4\% and do not exhibit any particular trend with time.

Finally, the PHOCAL response to the $^{241}$Am internal $\alpha$-source is displayed in Figure~\ref{fig:alpha} for the low and high gain signal acquisition mode. The response is normalised to the first data point and a constant pedestal determined at that point is subtracted. Figure~\ref{fig:alpha} shows a consistent down-drift for the two gain modes, that reach $-0.8$\% in the end of the period under analysis. Given the larger pedestal variation observed for the PHOCAL photodiode, seen in Figure~\ref{fig:pedestal}, the pedestal subtraction is made at a run-by-run level to correct for its effect. The correction has a substantial effect on the obtained photodiode response since the signal induced by the radioactive source (around 600 and 2500 ADC counts in low and high gain, respectively) is just three to six times larger than the pedestal values (around 100 and 750 ADC counts in low and high gain, respectively). Figure~\ref{fig:alpha} also shows the corrected responses, exhibiting a maximum relative variation of about 0.4\%.

The effects of fluctuations of the light monitoring system are taken into account in the calibration of the TileCal PMTs with a run-by-run correction factor. This will be described in Section~\ref{sec:calibration}.



\section{Calibration of the calorimeter with Laser II}
\label{sec:calibration}

\subsection{Calibration procedure}



As can be seen in Equation~\ref{eq:channelEnergy}, the reconstruction of the energy in TileCal depends on several constants, some of them being updated regularly. The main calibration of the TileCal energy scale is obtained using the caesium system~\cite{Blanchot:2020lyh}. However, since a caesium scan needs a pause in the $pp$ collisions of at least six hours, this calibration cannot be performed very often. Therefore, regular relative calibrations are accomplished between two caesium scans using the laser system. Moreover, during the LHC technical stop at the beginning of data taking period in 2016, few liquid traces coming from the caesium hydraulic system were found in the detector cavern. Since then until the end of Run~2, caesium scans were restricted to be taken only during the end of year technical stops, due to risk of the leak. In absence of the caesium calibration, the laser became the main calibration system, calibrating the PMTs and readout electronics. In order to address the fast drift of PMT response caused by the large instantaneous luminosity, the laser calibration constants were updated every 1--2 weeks, since July 2016. These constants were used in so-called prompt data processing, performed during the data taking period. 

Each year, the data recorded by the ATLAS detector is reprocessed. Data reprocessing consists of the update of the physics dataset (proton--proton and heavy ion collision runs) with updated conditions and calibration constants. Moreover, a reprocessing of the full Run~2 dataset was performed during LHC Long Shutdown 2 at the end of Run~2. This step is necessary to apply new reconstruction and calibration algorithms as well as the corrections that were impossible to be done or missed during prompt data processing. The IOVs are readjusted and chosen to coincide with the data taking periods. For laser calibration, they occur every 1--2 weeks in order to smoothly follow the evolution of PMTs response during the data taking period. 

The method to compute the laser constant $f_{\mathrm{Las}}$ introduced in Equation~\ref{eq:channelEnergy} is based on the analysis of specific laser calibration runs, taken daily during the data taking period, for which both the laser system photodiodes and the TileCal PMTs are read out. The laser calibration employs two types of successive laser runs:
\begin{itemize}
  \item Low Gain run (labeled as LG) consists of $\sim$10,000 pulses with a constant amplitude and the filter attenuation factor equal to 3,
  \item High Gain run (labeled as HG) consists of $\sim$20,000 pulses with a constant amplitude and the filter attenuation factor equal to 330.
\end{itemize}

The laser system is employed to perform the PMT response calibration relative to the previous global calibration of the TileCal detector with the caesium scan. Thus, to determine the laser calibration constants, a laser run taken close to the caesium scan is used to set the reference signals for each PMT. 
By definition, if the response of a channel to a given laser intensity is stable (the response of the PMT and of the associated readout electronics are stable), the laser constant $f_{\mathrm{Las}}$ is 1. The references were set close to the start of each year's $pp$ collision runs. 
The laser references and laser constants are stored in the conditions database. 

The laser calibration procedure evolved during Run~2. Due to increasing instantaneous luminosity and response variation observed in all PMTs, the methods to derive laser constants were adapted. The applied methods are described in detail in Section~\ref{sec:determination_of_the_calibration_constants}. 

\subsection{Determination of the calibration constants}
\label{sec:determination_of_the_calibration_constants}

The laser runs are constituted by a set of laser pulses with corresponding signal readout from the individual PMTs, from which the pedestal is subtracted. 
For each pulse, the normalised response of a PMT channel, the ratio $R_{i,p}$, is defined as:

\begin{equation}
    R_{i,p} = \frac{A_{i,p}}{A_{\mathrm{D6},{p}} }
    \label{eq:Rip}
\end{equation}

where $p$ denotes the pulse, $A_i$ is the reconstructed signal amplitude of the PMT readout channel $i$ and $A_{\mathrm{D6},{p}}$ is the signal amplitude measured by the photodiode 6 (D6) in the laser box. The D6 measures the laser light after the beam expander and probes the beam close to the TileCal PMTs in the best dynamic range among available photodiodes D6--D9. The average of the ratio $R_{i,p}$ over all pulses of the laser run, denoted as $R_i\equiv \langle R_{i,p}\rangle$, is analysed for each PMT.

The laser calibration factors employed to reconstruct the cell energy, in Equation~\ref{eq:channelEnergy}, are simply the relative response of the channel:

\begin{equation}
    f_{\mathrm{Las}}^i = \frac{R_i}{R_i^{\mathrm{ref}}}
    \label{eq:fLaser}
\end{equation}

where $R_i^{\mathrm{ref}}$ is the normalised response of the PMT channel $i$ during the laser reference run. For monitoring purposes, these factors are usualy presented in percentage as a relative response variation:

\begin{equation}
    (f_{\mathrm{Las}}^i - 1)\times 100\;[\%]
\label{eq:PMTdriftCorrected}
\end{equation}

The measurement of $f_{\mathrm{Las}}^i$ may be influenced by instabilities with origin at the laser system itself, both at a global level, i.e. affecting equally all the detector PMTs, or at the fibre level, i.e. affecting the set of PMTs associated with each clear fibre. To take these effects into account, global and fibre corrections are determined, such that the corrected laser constant reads as:

\begin{equation}
    f_{\mathrm{Las}}^i \to f_{\mathrm{Las}}^i \times \frac{1}{\alpha_{\mathrm{G}} \times \alpha_{\mathrm{f}(i)}}
    \label{eq:fLaserOpticsCorrections}
\end{equation}

\begin{itemize}                                                                               
\item The global correction $\alpha_{\mathrm{G}}$ is associated with a coherent drift of all channels. The effect can be related to an instability of the reference diode, from the variation of light received by the TileCal PMTs or common ageing of the long fibres.
                                                                                       
\item The fibre correction $\alpha_{\mathrm{f}(i)}$, computed per fibre $\mathrm{f}(i)$, is associated with a time variation of the light transmission from fibre to fibre.
\end{itemize}                                                                                

During Run~2, two methods were used to evaluate these optics corrections: the so-called Direct and Combined methods. In the Direct method, the global and fibre corrections are simply determined from the average response variations of a set of stable PMTs reading outermost and least irradiated cells in the D layer used as references, and the sub-set of D-layer PMTs associated with the fibre, respectively. 
This method was used to calibrate and monitor the detector response during 2015--2017 data taking but revealed to be inadequate for calibration when the response of the reference PMTs started to vary due to larger integrated currents in the middle of the 2017 run. Then the Combined method was developed and employed in the 2018 TileCal calibration and also for the reprocessing of 2017 data. Instead of relying on the stability of a set of reference PMTs, the gain of the PMT is explicitly evaluated to determine the optics corrections by the Combined method.

\subsubsection*{Direct method}

In the Direct method, the global correction is evaluated from the relative response of all PMTs reading cells in the D-layer:

\begin{equation}
  \alpha_{\mathrm{G}} = \bigg\langle \frac{R_i}{R_i^{\mathrm{ref}}} \bigg\rangle^{\mathrm{D-cells}}
  \label{eq:globalCorrectionDM}
\end{equation}

The fibre corrections are evaluated using information from PMTs of the D layer for the fibres associated to the LB, and from PMTs reading the D, B13, B14 and B15 cells for the EB\footnote{These cells are less exposed to particle fluence, so their readout PMTs experience smaller integrated currents and a more stable response.}, corrected from global effects. This quantity is evaluated for each long clear fibre $\mathrm{f}(i)$ as 

\begin{equation}
  \alpha_{\mathrm{f}(i)} = \frac{1}{\alpha_{\mathrm{G}}} \bigg\langle \frac{R_i}{R_i^{\mathrm{ref}}} \bigg\rangle^{\mathrm{D,B-cells}}_{\mathrm{f}(i)}
  \label{eq:fibreCorrectionDM}
\end{equation}

In Equations~\ref{eq:globalCorrectionDM} and~\ref{eq:fibreCorrectionDM}, $\langle\;\rangle$ represents a geometric weighted average, where the weight is proportional to the number of laser pulses in the run, and the average RMS of the PMT signals.

Saturated channels, channels with bad status in the TileCal condition database, and channels for which the absolute difference between the applied and requested HV is above 
10~V ($\Delta \mathrm{HV}>$10~V) are excluded from the computation of the optics corrections. These represent less than 2\% of total number of channels in TileCal. Moreover, an iterative procedure rejects outlier channels, more than $3\sigma$ apart from the $R_i/R_i^{\mathrm{ref}}$ distribution average.

\subsubsection*{Combined method}

In the Combined method, the actual PMT gain is measured based on the statistical nature of photoelectron production and multiplication inside the PMT. It assumes that the noise is negligible with respect to the laser-induced PMT signals~\footnote{The signal-to-noise ratio for the laser signal is of the order of 300.} and that the laser light is coherent. Under these conditions, the two main contributions to the PMT signal fluctuations to the laser scans are the poissonian fluctuations in the photoelectron emission spectrum and multiplication, and the variation of the intensity of the light source~\cite{Bures:74}. The PMT gain $G$ can be written as:

\begin{equation}
	G = \frac{1}{f \cdot e}\cdot \left( \frac{\mathrm{Var}[q]}{\langle q\rangle} - k \cdot \langle q\rangle \right)
	\label{eq:gain_statistical_method}
\end{equation}

where $e=1.6\times 10^{-19}$~C is the electron charge constant, $f$ stands for the excess noise factor extracted from the known gain of the individual PMT dynodes~\cite{1073137}. For the eight dynode TileCal PMTs, $f=1.3$ at the nominal gain of $G=10^5$, $\langle q \rangle$ is the average value of PMT anode charge associated to each laser pulse, and $\mathrm{Var}[q]$ is the variance of the anode charge distribution. 
The coherence factor $k = \frac{\mathrm{Var}[I]}{\langle I\rangle ^2}$ depends on the characteristics of the light source itself~\footnote{$⟨I⟩$ ($\mathrm{Var}[I]$) is the average (variance) light intensity $I$ of a pulsed source on the PMT cathode.} but not on the light intensity. The factor $k$ ranges from 0, for an ideal fully coherent light source, to 1, for a totally incoherent light source, and is determined with a set of PMTs measuring the same light source. For any PMT pair $i$ and $j$, $k$ is given by the average PMT measured charge $q_i$ and $q_j$ respectively, and the covariance $\mathrm{Cov}[q_i, q_j]$ of the charge measurements, as:

\begin{equation}
	k = \frac{\mathrm{Cov}[q_i, q_j]}{\langle q_i\rangle \langle q_j\rangle}
	\label{eq:PMT_excess_noise_factor}
\end{equation}

In order to decrease the dependence of the gain measurement on the $k$ factor determination, to which the sensitivity is more limited, the PMT gain is analysed in high gain laser calibration runs taken with filter wheel in position~8 with 31.6\% transmission (optical density of 2.5). For these runs the light intensity is lower leading also to a lower average PMT anode charge $\langle q \rangle$, thus the $k$ term in Equation~\ref{eq:gain_statistical_method} has a smaller effect on the gain measurement.

Moreover, since the PMT gain determination presents significant fluctuations, the average over a set of runs within $\pm$10 days around the laser reference run is taken to set the PMT reference gain, $G_i^{\mathrm{ref}}$. The PMT gain $G_i$ is used as an independent measure of the PMT signal and as the basis to evaluate the optics corrections by the Combined method. The global correction is determined from the average ratio between the PMT relative response and the PMT relative gain using PMTs reading the D-layer and the BC1, BC2, B13, B14, B15 cells:

\begin{equation}
    \alpha_{\mathrm{G}} = \bigg\langle \frac{R_i}{R_i^{\mathrm{ref}}} \Big/ \frac{G_i}{G_i^{\mathrm{ref}}} \bigg\rangle^{\mathrm{D,B-cells}}
    \label{eq:global_combined}
\end{equation}

The fibre corrections are determined in approximately the same way as the global correction except that the average runs over all the channels connected to a common long fibre $\mathrm{f}(i)$, with the global correction taken into account to avoid double correcting:

\begin{equation}
    \alpha_{\mathrm{f}(i)} = \frac{1}{\alpha_{\mathrm{G}}} \bigg\langle \frac{R_i}{R_i^{\mathrm{ref}}} \Big/ \frac{G_i}{G_i^{\mathrm{ref}}} \bigg\rangle_{\mathrm{f}(i)}
    \label{eq:fibre_combined}
\end{equation}

As for the Direct method, the PMTs having bad status or with $\Delta \mathrm{HV}>$10~V or saturated channels are discarded from analysis.

\subsection{Evolution of the optics corrections}

\begin{figure}[htbp]
\centering
\subfloat[\label{fig:optics_correction_2018_a}]{\includegraphics[width=0.5\linewidth]{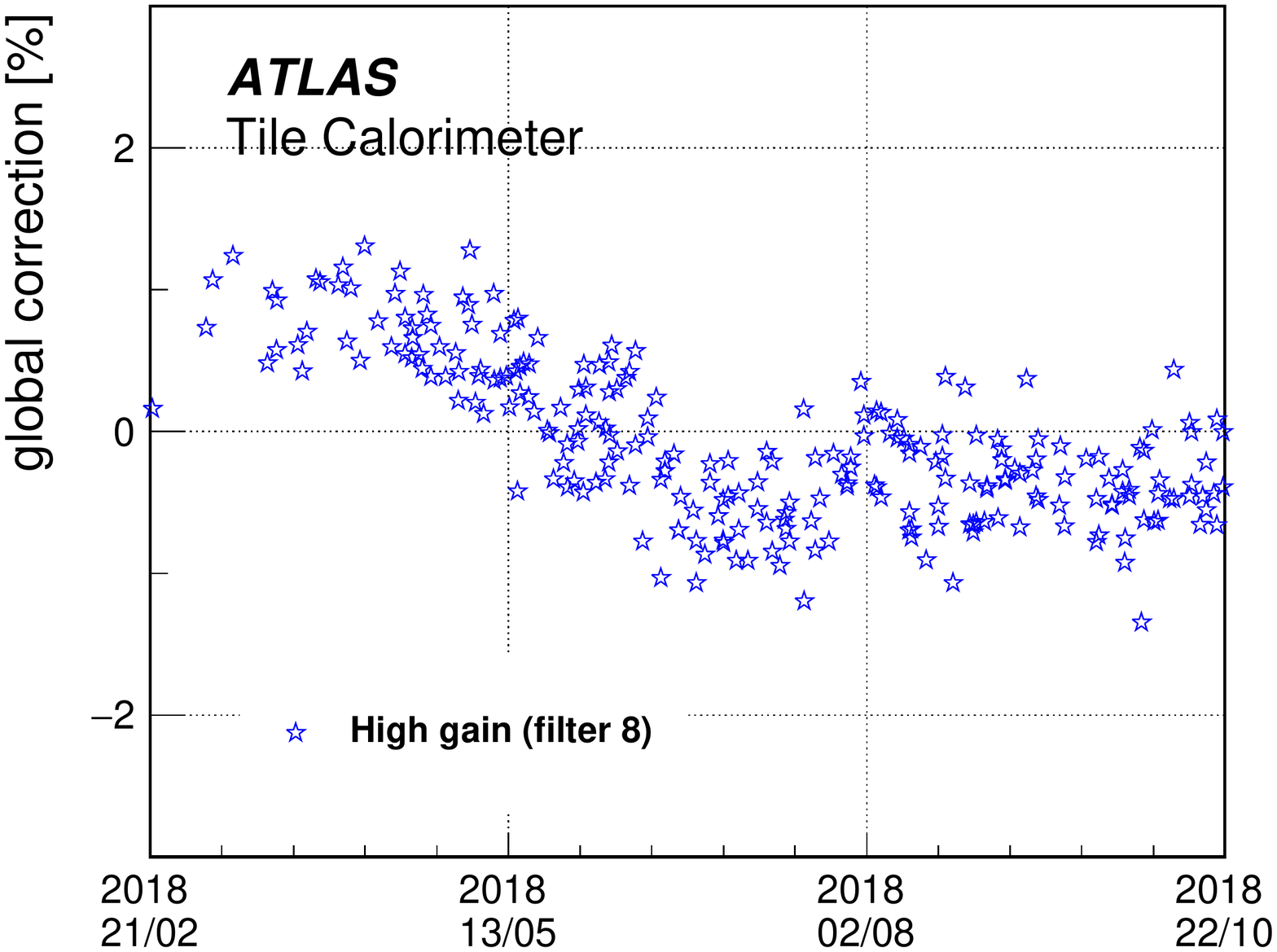}}
\subfloat[\label{fig:optics_correction_2018_b}]{\includegraphics[width=0.5\linewidth]{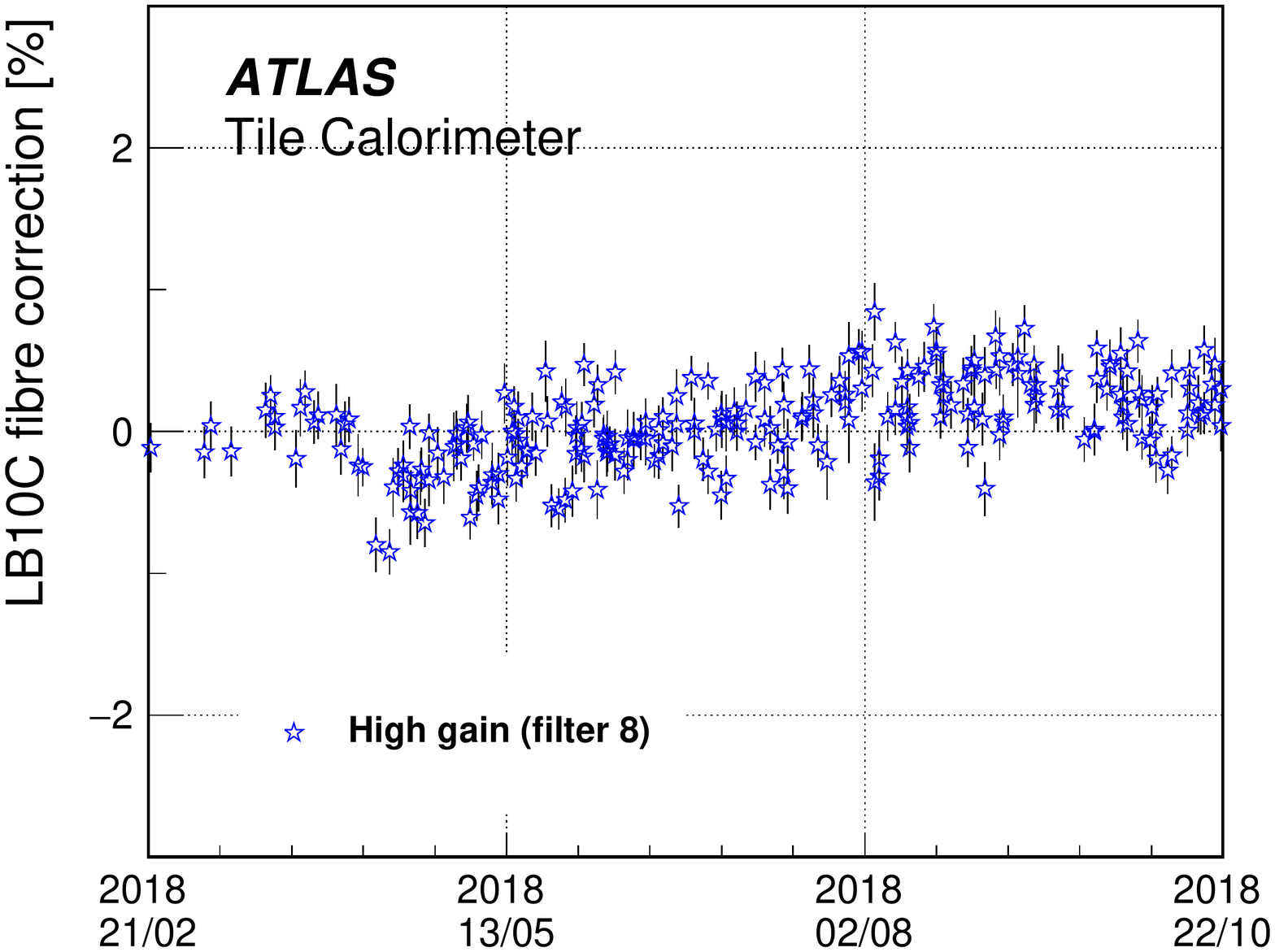}}
\caption{Evolution of the (a) global correction and (b) LB10C fibre correction associated with even/odd numbered PMTs in LBA10/LBC10 over time in 2018. The corrections are determined using laser high gain runs with the Combined method and are calculated as a weighted geometric mean. The corresponding errors are included in the data points.}
\label{fig:optics_correction_2018}
\end{figure}

Figure~\ref{fig:optics_correction_2018} shows the time evolution in 2018 of the global correction and the LB10C fibre correction (associated with the even/odd numbered PMTs in LBA10/LBC10), both shown in percentage and determined with the Combined method using laser runs taken in high gain. The global correction in 2018 is stable in time within 1\% and the correction is about the same order. During Run 2, the magnitude of this correction did not exceed 2.5\%. The fibre correction shown is generally representative of the 384 clear fibres in total. For all the years, the magnitude of the corrections did not exceed 1\% and was also found to be constant throughout the time.

The global correction dominates the scale of the PMT calibration. Its precision should match the global scale uncertainty on the PMT calibration assessed from laser and caesium comparisons presented in Section~\ref{sec:CsLas}, and thus be better than 0.4\%. The accuracy on the global correction was further assessed using two symmetric sets of PMTs, one composed of PMTs reading the TileCal A side and another with PMTs installed in the C side to derive independent corrections. The corrections obtained for the A and C sides matched well below the sub-percent level for all years in Run~2, attesting the robustness of the Combined method at disentangling the effects of fluctuations in the monitored light intensity common to all PMTs.

\subsection{Comparison with caesium calibration}
\label{sec:CsLas}

The response variation of PMTs measured with the laser system should match the full detector response variation obtained with the caesium system within short periods of time, where fluctuations from the scintillators and WLS fibers can be safely neglected. Thus, the comparison between the laser and caesium measurements constitutes a procedure to validate the laser algorithm itself, employed to validate the Combined method.

\begin{figure}[htbp]
\centering
\subfloat[\label{fig:CsLas2015Nov_a}]{\includegraphics[width=0.51\linewidth]{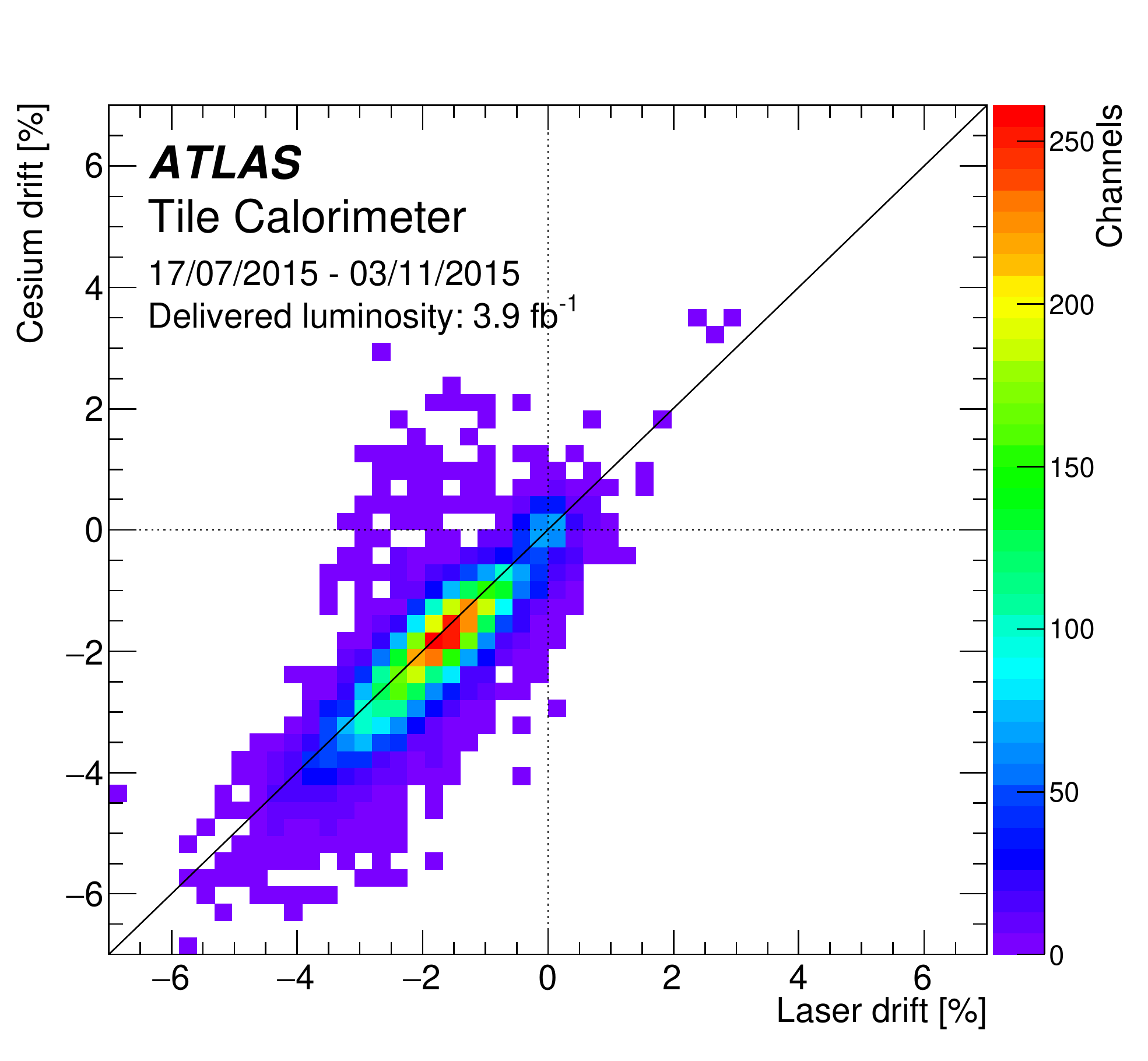}}
\subfloat[\label{fig:CsLas2015Nov_b}]{\includegraphics[width=0.49\linewidth]{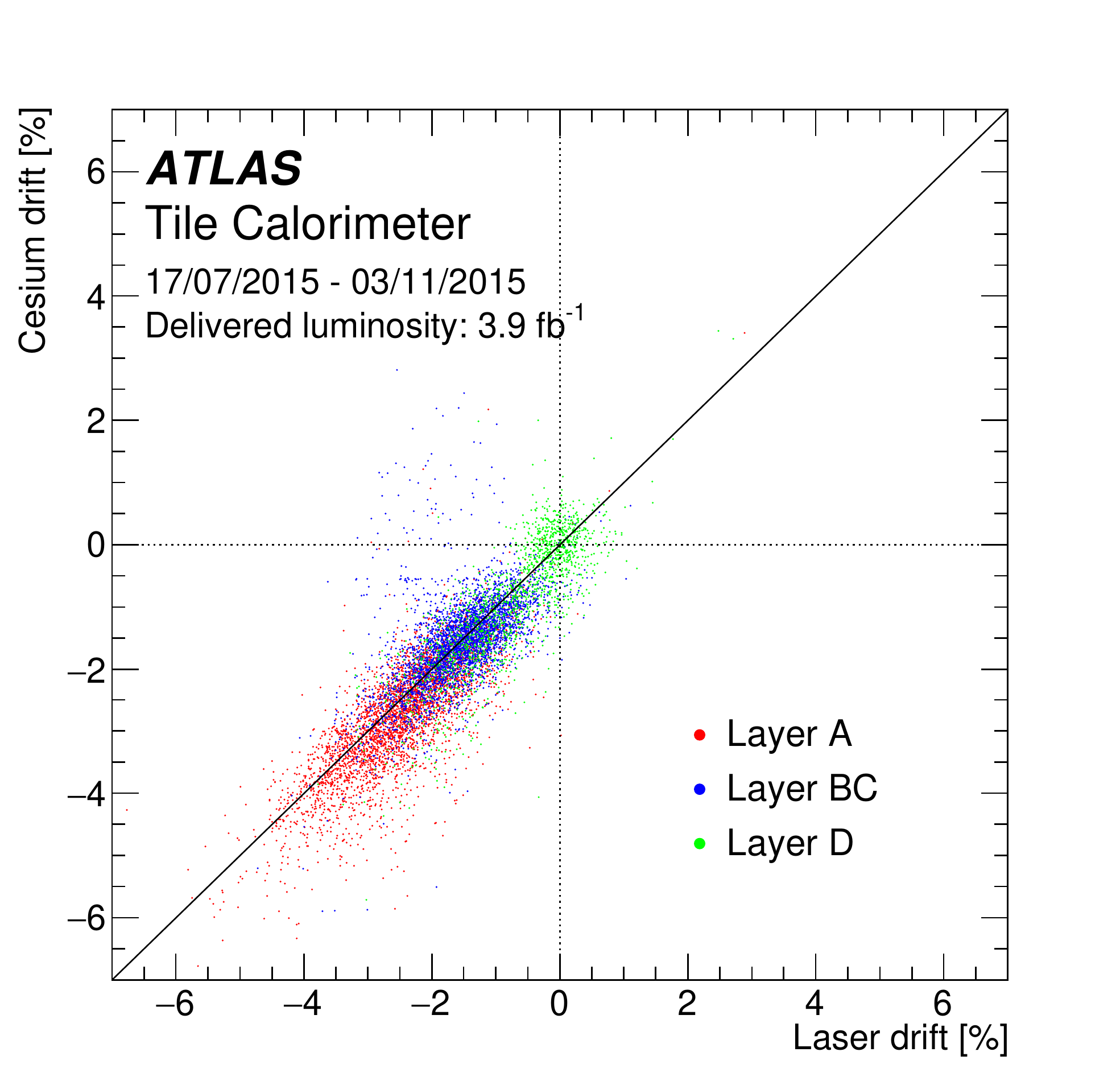}}
\caption{Response variation (in \%) measured by caesium (y-axis) and by laser employing the Combined method (x-axis) between July 17 and November 3, 2015 for (a) all TileCal channels (9852) and (b) the channels in the A-, B/BC- and D-layers. Special channels not calibrated by the caesium system, such as the E-cells, are not included.
}
\label{fig:CsLas2015Nov}
\end{figure}

During 2015 and 2016, three periods of low integrated luminosity were available within consecutive caesium scans. Figure~\ref{fig:CsLas2015Nov} shows the response variation between July 17 and November 3, 2015, obtained with the caesium system as a function of the response variation obtained with the laser. The results are displayed at channel level and separating channels per layer A, B/BC and D. The great majority of channels have the same response variation for laser and for caesium.

\begin{figure}[htbp]
\centering
\subfloat[\label{fig:CsLas2015Nov_1d_a}]{\includegraphics[width=0.5\linewidth]{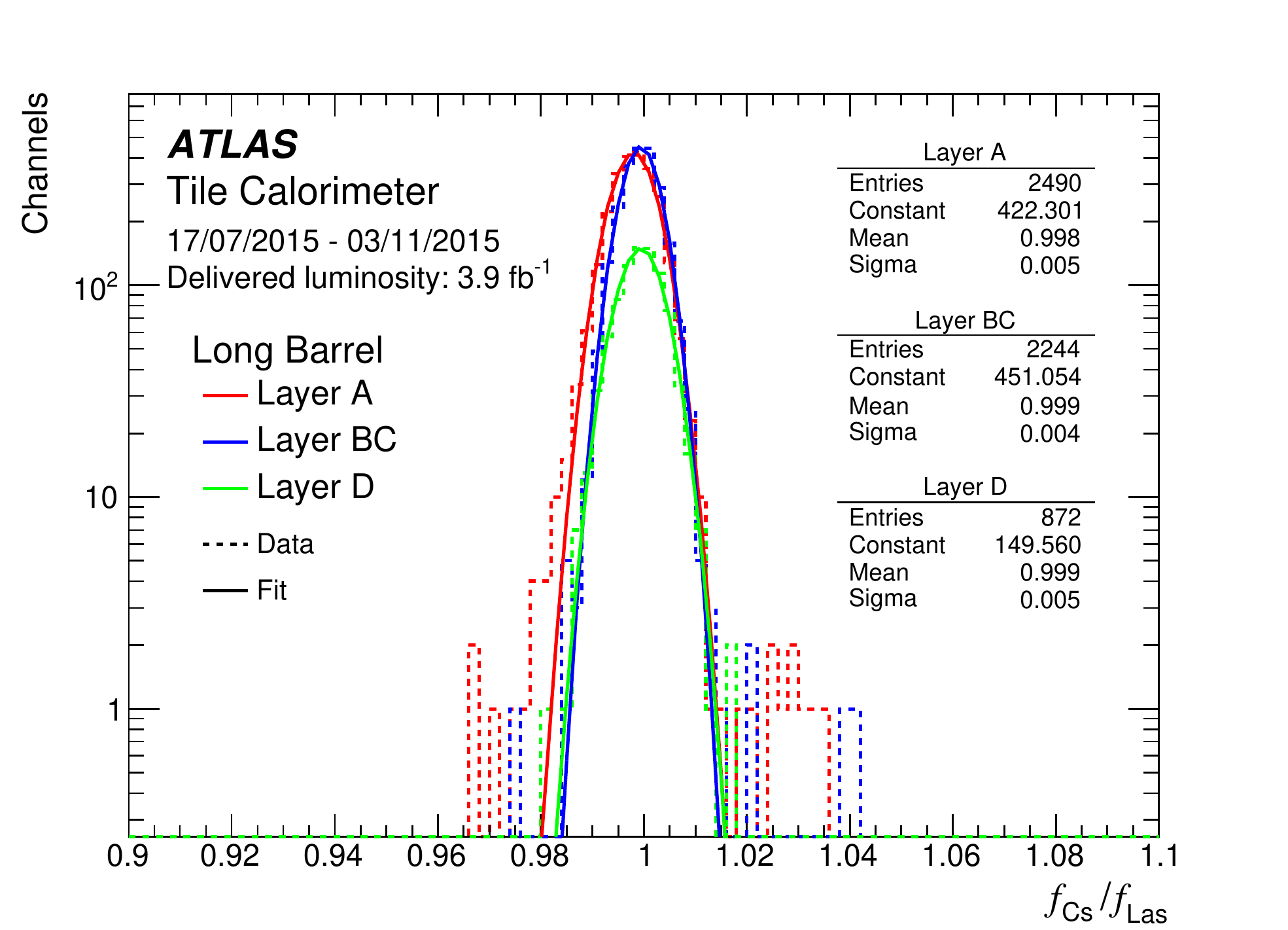}}
\subfloat[\label{fig:CsLas2015Nov_1d_b}]{\includegraphics[width=0.5\linewidth]{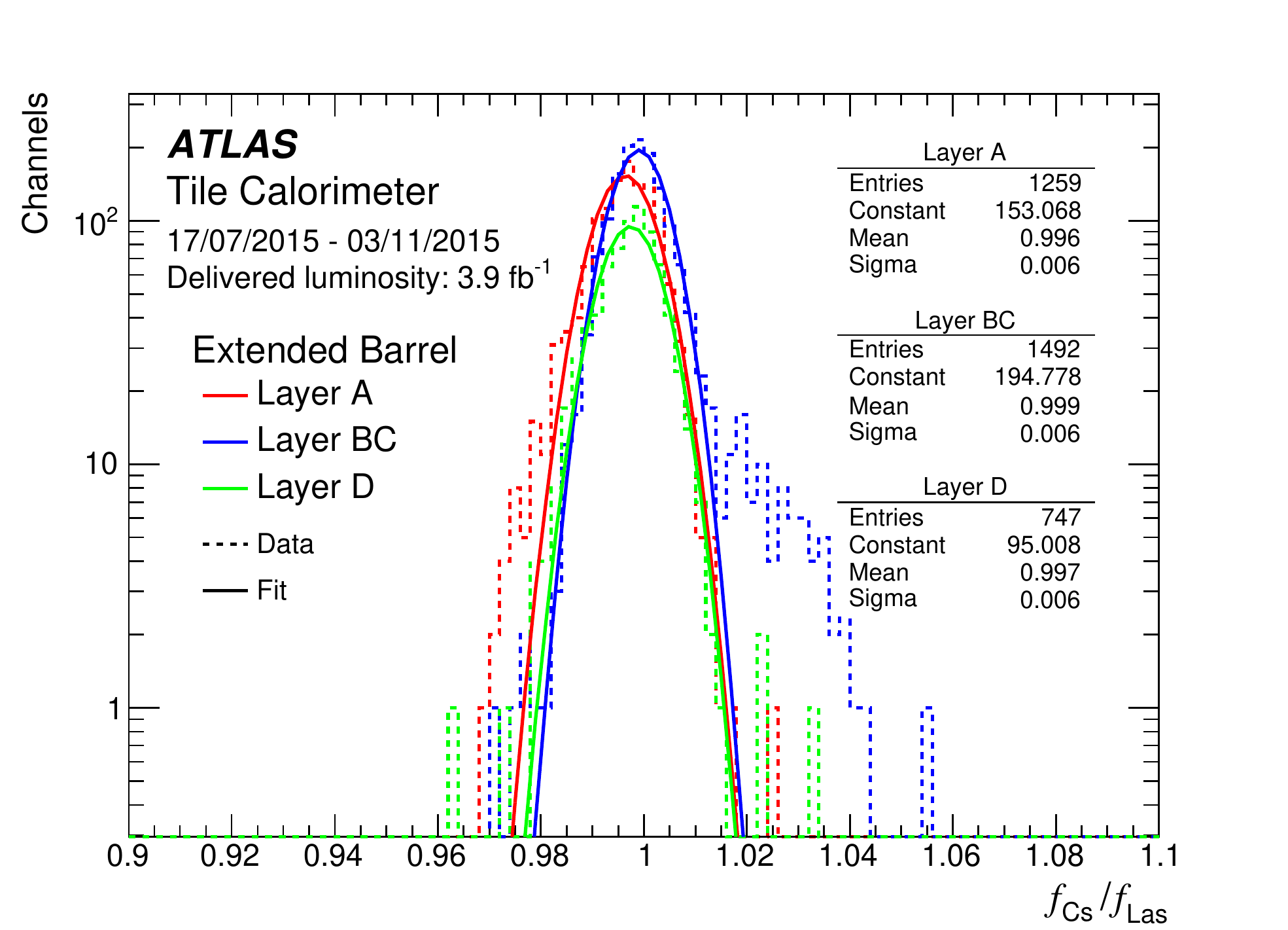}}
\caption{Ratio between the caesium calibration constants ($f_\mathrm{Cs}$) and the laser calibration constants calculated with Combined method ($f_\mathrm{Las}$) for channels in Layer A, B/BC and D in the (a) Long Barrel and (b) Extended Barrel. Special channels not calibrated by the caesium system, such as the E-cells, are not included.}
\label{fig:CsLas2015Nov_1d}
\end{figure}

The corresponding distribution of the ratio between the caesium constants ($f_\mathrm{Cs}$) and the laser calibration constants ($f_\mathrm{Las}$), calculated to address the response variation of the PMTs during the same period of time, 
is shown in Figure~\ref{fig:CsLas2015Nov_1d} 
separated by layer and Long/Extended barrel. 
Each distribution is fitted with a Gaussian function to measure its average and standard deviation. The differences observed between the caesium and the laser systems are more evident in the extended barrel and on the A layer. These regions of the calorimeter are less shielded and thus the effects of radiation damage to scintillator and WLS fibre are faster. The average difference is well below 0.1\% and the standard deviation is 0.6\%.

For the three periods analysed, the maximum average difference observed was 0.4\%. This value is taken as the uncertainty on the scale of the PMT calibration with laser, an improvement over the corresponding systematic uncertainty of 0.4 to 0.6~\% found in Run~1~\cite{bib:laser_run_1}.

\subsection{Uncertainties on the PMT calibration}
\label{sec:uncertainty}

\begin{figure}[htbp]
\centering
\includegraphics[width=0.5\textwidth]{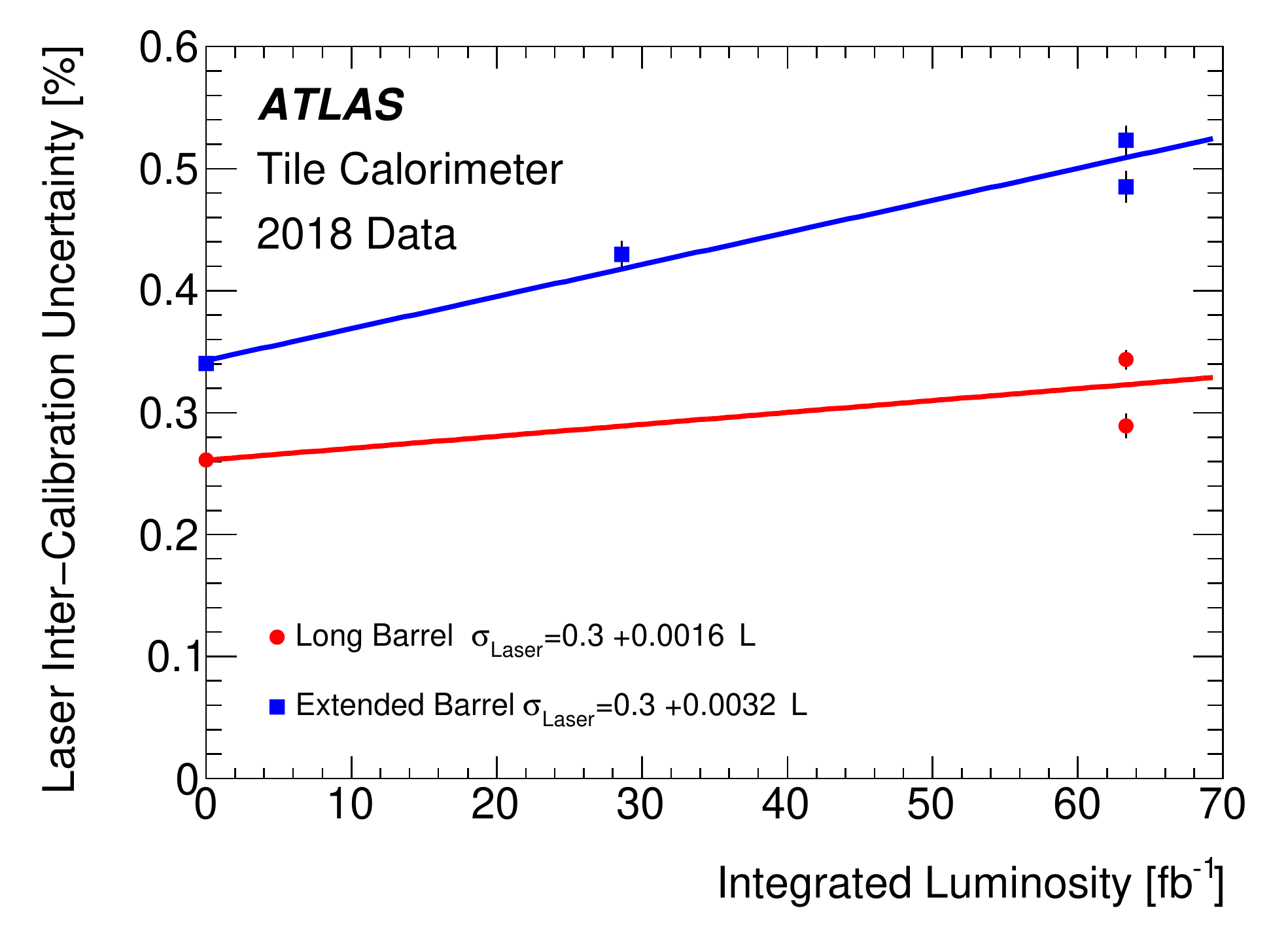}
\caption{Uncertainties on the PMT inter-calibration with the Laser~II system using the Combined method as a function of the integrated luminosity for the Long Barrel and for the Extended Barrel. The results are obtained using laser and caesium calibration data collected in 2018. The two points at 63.3~fb$^{-1}$ result from two successive caesium scans without LHC beam. The uncertainty is parametrised as a function of the luminosity by fitting the data points with a linear function. A global scale systematic resulting from direct comparison between laser and caesium data was found to be 0.4\%. This value should be summed in quadrature to obtain the total laser uncertainty. In 2018, three caesium scans were performed in LB (red points) and four in EB (blue points).}
\label{fig:sigma_Las}
\end{figure}

Besides the systematic uncertainty on the PMT calibration scale, the uncertainty on the PMT relative inter-calibration, mostly sourced at the fibre correction procedure and at the channel-level readout, is evaluated. To do so, an indirect comparison between the responses to caesium source and laser, measured with left and right PMTs reading the same cell, is performed evaluating the following observable: 

\begin{equation}
\Delta f^{\mathrm{L-R}}_{\mathrm{Cs/Las}}=\left( \frac{f^{\mathrm{L}}_{\mathrm{Cs}}}{f^{\mathrm{L}}_{\mathrm{Las}}}- 
\frac{f^{\mathrm{R}}_{\mathrm{Cs}}}{f^{\mathrm{R}}_{\mathrm{Las}}} \right)
\label{eq:sys_Cs-Laser}
\end{equation}

where $f^{\mathrm{L(R)}}_{\mathrm{Las}}$ and $f^{\mathrm{L(R)}}_{\mathrm{Cs}}$ are the calibration constants corresponding to the cell relative response to laser and caesium source measured by the left (right) channel. With this quantity, the scintillator effects common to both left/right readouts are cancelled out. Assuming that the WLS fibre response from the left and right sides of the cell has a similar behaviour, the width of the distribution of $\Delta f^{\mathrm{L-R}}_{\mathrm{Cs/Las}}$ is driven by the uncertainties of the laser measurement and caesium measurements. The inter-calibration systematic uncertainty on the laser calibration was then determined by disentangling the contributions from the caesium uncertainty and constraining with measurements of $f^{\mathrm{L}}_{\mathrm{Cs}}-f^{\mathrm{R}}_{\mathrm{Cs}}$ and $f^{\mathrm{L}}_{\mathrm{Las}}-f^{\mathrm{R}}_{\mathrm{Las}}$. The results obtained with 2018 data are shown in Figure~\ref{fig:sigma_Las}. A dependence of the systematic uncertainty on the integrated luminosity $L$, more pronounced for the extended barrel, is observed. The effect is due to a correlation between the integrated PMT charge and the response down-drift, with consequent increase in the spread of the response for a given PMT sample.

The total uncertainty on the laser calibration of a PMT, corresponding to the quadratic sum of the 0.4\% scale systematic and the luminosity $L$-dependent inter-calibration uncertainty ($L$ in fb$^{-1}$), in the Long Barrel ($\sigma_{\mathrm{Las,tot}}^{\mathrm{LB}}$) and in the Extended Barrel ($\sigma_{\mathrm{Las,tot}}^{\mathrm{EB}}$) yields:

\begin{equation}
\begin{aligned}
\sigma_{\mathrm{Las,tot}}^{LB} [\%] =0.4\oplus(0.3+0.0016\times L)\; [\%] \\
\sigma_{\mathrm{Las,tot}}^{EB} [\%] =0.4\oplus(0.3+0.0032\times L)\; [\%]
\end{aligned}
\end{equation}

\subsection{Overview of the PMT response}
The laser system is used to measure the evolution of the PMT response as a function of time. 
The Combined method, discussed in Section~\ref{sec:determination_of_the_calibration_constants}, is utilised to calculate the response variation with respect to a set of reference runs. In particular, the Equation~\ref{eq:PMTdriftCorrected} is used to obtain the response variation for each PMT. 
Channels marked with bad data quality status, unstable high voltage or flagged as problematic by any calibration system are discarded. For each cell type, the average response is obtained by a Gaussian fit to the distribution of PMT response variation. The $\chi ^2$ fit method is applied. The Gaussian approximation is used in order to obtain the average variation that is not affected by outliers. 

\begin{figure}[t]
\centering
    \includegraphics[width=1.0\textwidth]{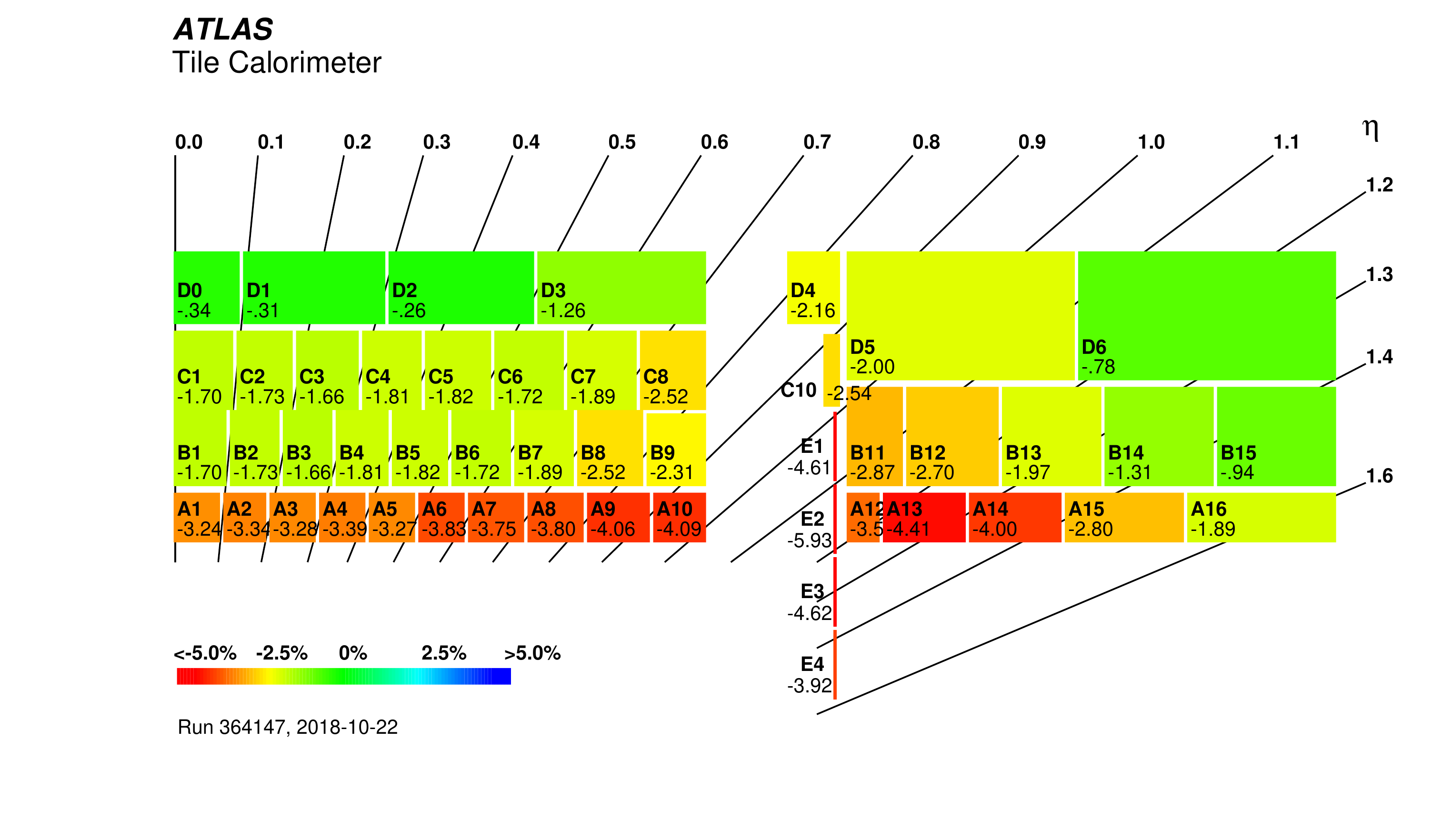}
    \caption{The mean response variation in the PMTs for each cell type, averaged over $\phi$, observed during the entire $pp$ collisions data-taking period in 2018 (between laser calibration runs taken on 18 April 2018 and 22 October 2018) calculated using the Combined method. For each cell type, the response variation is defined as the mean of a Gaussian fit to the response variations in the channels associated with given cell type. A total of 64 modules in $\phi$ were used for each cell type, with the exclusion of known pathological channels.
    }\label{fig:map_2018}
\end{figure}

A sample of the mean response variation in the PMTs for each cell type averaged over $\phi$, measured with the laser system during the entire $pp$ collisions data-taking period in 2018, is shown in Figure~\ref{fig:map_2018}. The most affected cells are those located at the inner radius and in the gap and crack region with down-drift up to 4.5\% and 6\%, respectively. Those cells are the most irradiated and their readout PMTs experience the largest anode current. 

\begin{figure}[t]
\centering
    \subfloat[\label{fig:phimap_2018_a}]{\includegraphics[width=0.495\linewidth]{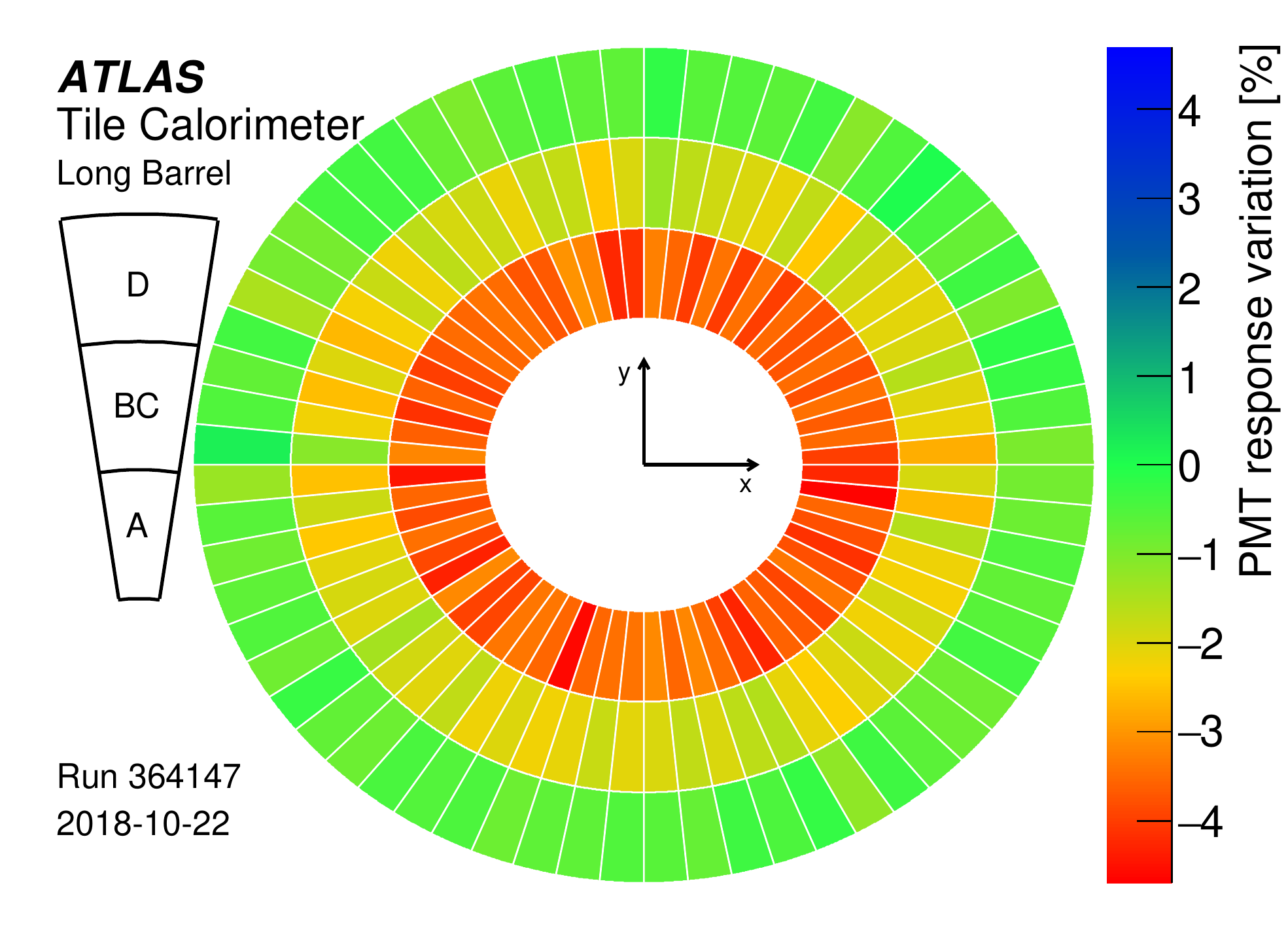}}\hfill
    \subfloat[\label{fig:phimap_2018_b}]{\includegraphics[width=0.495\linewidth]{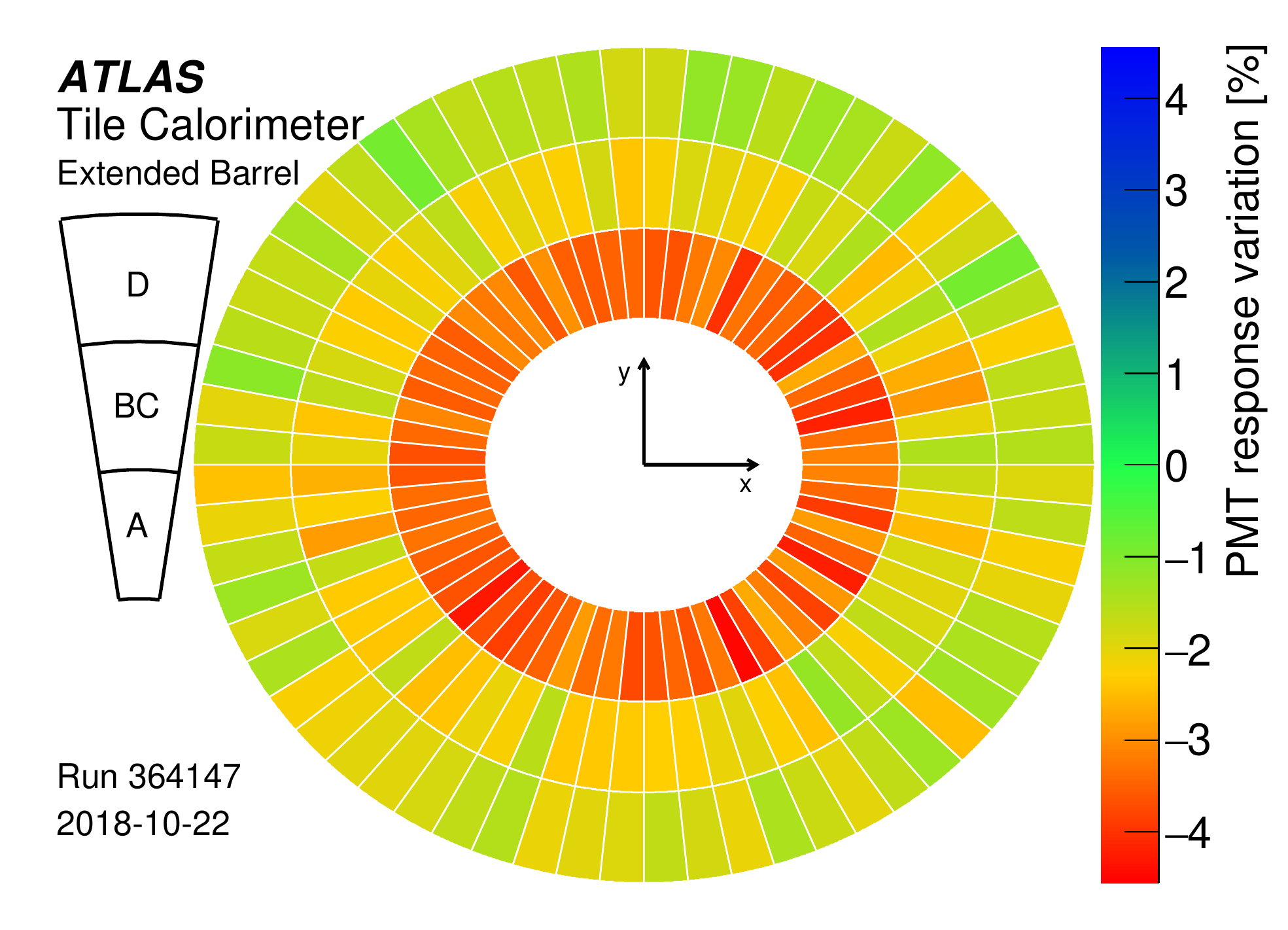}}
    \caption{The mean response variation in the PMTs for each cell type, averaged over $\eta$, observed during the entire $pp$ collisions data-taking period in 2018 (between laser calibration runs taken on 18 April 2018 and 22 October 2018) in LB (a) and EB (b), calculated using the Combined method. For each cell type, the response variation is defined as the mean of a Gaussian fit to the response variations in the channels associated with given cell type. Known pathological channels were excluded.
    }\label{fig:phimap_2018}
\end{figure}

Figure~\ref{fig:phimap_2018} shows the average response variation of the channels per layer and along the azimuthal angle $\phi$ for the same period in 2018. Each $\phi$ bin corresponds to one LB/EB module averaged over the A and C sides. Channels with low signal amplitude, bad data quality status or unstable high voltage are discarded in the average response calculation. It can be seen that PMTs reading the cells in layers closest to the beam axis, composed of A cells, are the most affected. Next layers, formed of the BC and D cells are significantly less affected. We observe larger uniformity across the modules in $\phi$ in layers with a larger number of channels (eg. 40 channels in the LB A layer, see Figure~\ref{fig:map_2018}), where the effect of discarding one bad quality channel has less impact. On the other hand, layers for which the spread in the response of the channels is larger (eg. EB layer A against LB layer A, see Figure~\ref{fig:map_2018}) are more affected in the $\phi$ uniformity with bad channel removal.

\begin{figure}[t]
\centering
    \subfloat[\label{fig:LaserDrift_run2_a}]{\includegraphics[width=0.5\linewidth]{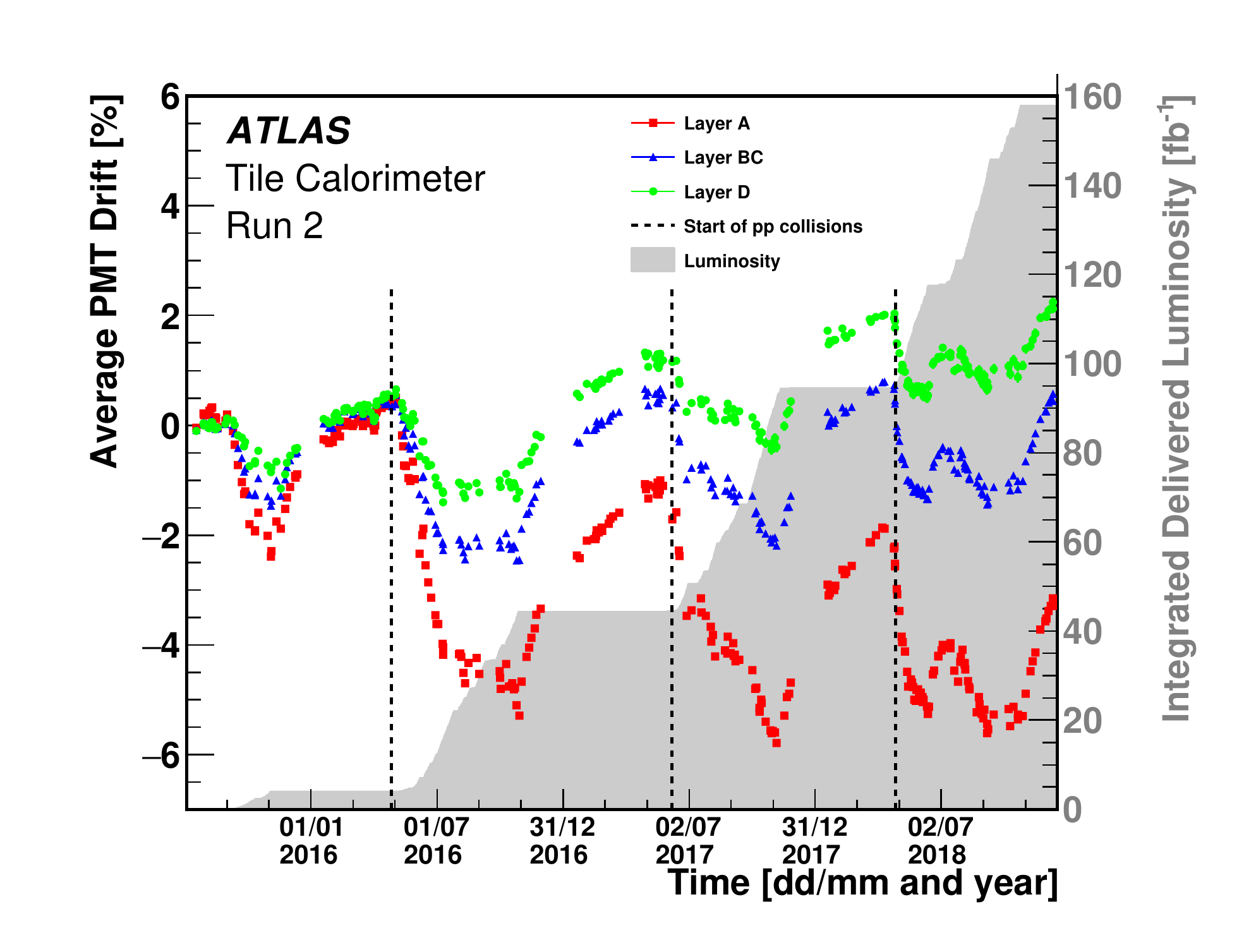}}
    \subfloat[\label{fig:LaserDrift_run2_b}]{\includegraphics[width=0.5\linewidth]{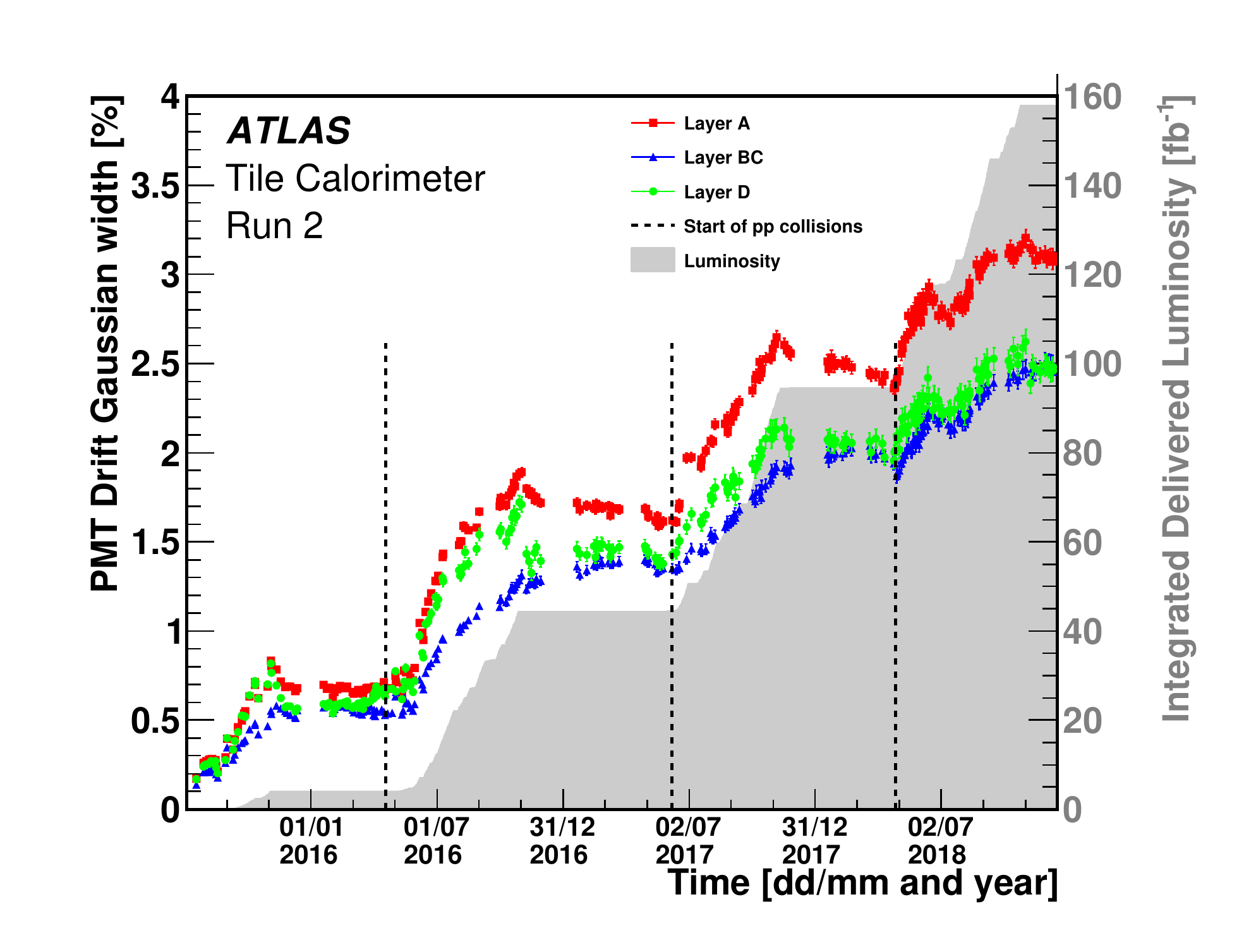}}
    \caption{The mean response variation in the PMTs (a) and Gaussian width (b) for each layer, as a function of time, observed during the entire Run~2 (between stand-alone laser calibration runs taken on 17 July 2015 and 22 October 2018). For each layer, the response variation is defined as the mean of a Gaussian fit to the variations in the channels associated with given layer. Known pathological channels are excluded. The laser calibration runs were not taken during the ATLAS end-of-year technical stops. Moreover, the laser system was not operational due to technical problems in the period September 10--27, 2016. Thus, no laser data can be seen in the plots for these time intervals. The LHC delivered luminosity is shown for comparison in grey. The vertical dashed lines show the start of $pp$ collisions in respective years.
    }\label{fig:LaserDrift_run2}
\end{figure}

Figure~\ref{fig:LaserDrift_run2_a} shows the time evolution of the mean response variation in the PMTs for each layer observed during the entire Run~2. The PMT response variation strongly depends on the delivered luminosity by the LHC. Therefore, the delivered luminosity is also shown for comparison. The observed PMTs response variation is the result of three competing factors: i) the constant up-drift observed when PMTs are in rest; ii) the down-drift during high instantaneous luminosity periods when PMTs are under stress; iii) the fast partial recovery after stress observed during technical stops. These effects result in -6\% accumulated mean response variation in the PMTs for the cells located at the inner layer at the end of Run~2. For the B/BC and D layers, the average PMT response degradation during $pp$ collisions was almost totally recovered in technical stops, resulting in $-1.5$\% accumulated PMT response variation at the end of Run~2 for the layer~B/BC and even in +0.5\% balance for the layer~D. 

Figure~\ref{fig:LaserDrift_run2_b} shows the Gaussian width distribution as a function of time observed for each layer during the entire Run~2. The Gaussian width for all layers increases with time during high instantaneous luminosity periods when PMTs are under stress. It is caused by the different behaviour of different PMTs over time which are at different $|\eta|$ positions. During technical stops, when PMTs are at rest, some inversion of this effect is observed resulting from the recovery of the most affected PMTs to the average response in a given layer or cell type.

\FloatBarrier


\section{Monitoring with Laser during physics runs}
\label{sec:laseringap}

\subsection{Time monitoring}

\newcommand{\tlaser}{$t^{\mathrm{laser}}_{\mathrm{chan}}$}

The TileCal does not only provide a measurement of the energy that is
deposited in the calorimeter, but it also measures the time when
particles and jets hit the calorimeter cell. This information is
particularly utilised in the removal of signals which do not originate
from $pp$ collisions along with the time-of-flight
measurements of hypothetical heavy slow particles that would reach the
calorimeter.

The time calibration is also important for the energy reconstruction
itself. As explained in Section~\ref{sec:signal_energy_reco},
physics collision events are reconstructed with the OF algorithm (see
Eq.~(\ref{eq:of})), whose weights depend on the 
expected phase. If the real signal phase significantly differs from the
expected one, the reconstructed amplitude is underestimated. 
Consequently, the time synchronisation of all calorimeter channels
represents an important issue. While the final time calibration is performed
with $pp$ collision data, laser data are extensively used to check its
stability and to spot eventual problems.

Laser calibration events are shot during empty bunch-crossings of
physics runs with a frequency of about 1~Hz. These events, also
referred to as laser-in-gap events, were originally proposed for the
PMT response monitoring. However, they are also extensively used for the
time calibration stability monitoring.

The monitoring tool creates a 2D histogram for each channel and fills
them with the reconstructed time ({\tlaser}) and luminosity block~\footnote{The luminosity block is the data elementary unit within a run, and corresponds to up to one minute of collision data for which the detector conditions or software calibrations remain approximately
constant.} for each
event. These histograms are stored and automatically examined for
anomalies, which include average {\tlaser} being off zero in at least
few consecutive luminosity blocks, unstable {\tlaser} or fraction of
events off zero by more than 20~ns.

\begin{figure}[htbp]
\centering
    \subfloat[\label{fig:timing_jump_a}]{\includegraphics[width=0.5\linewidth]{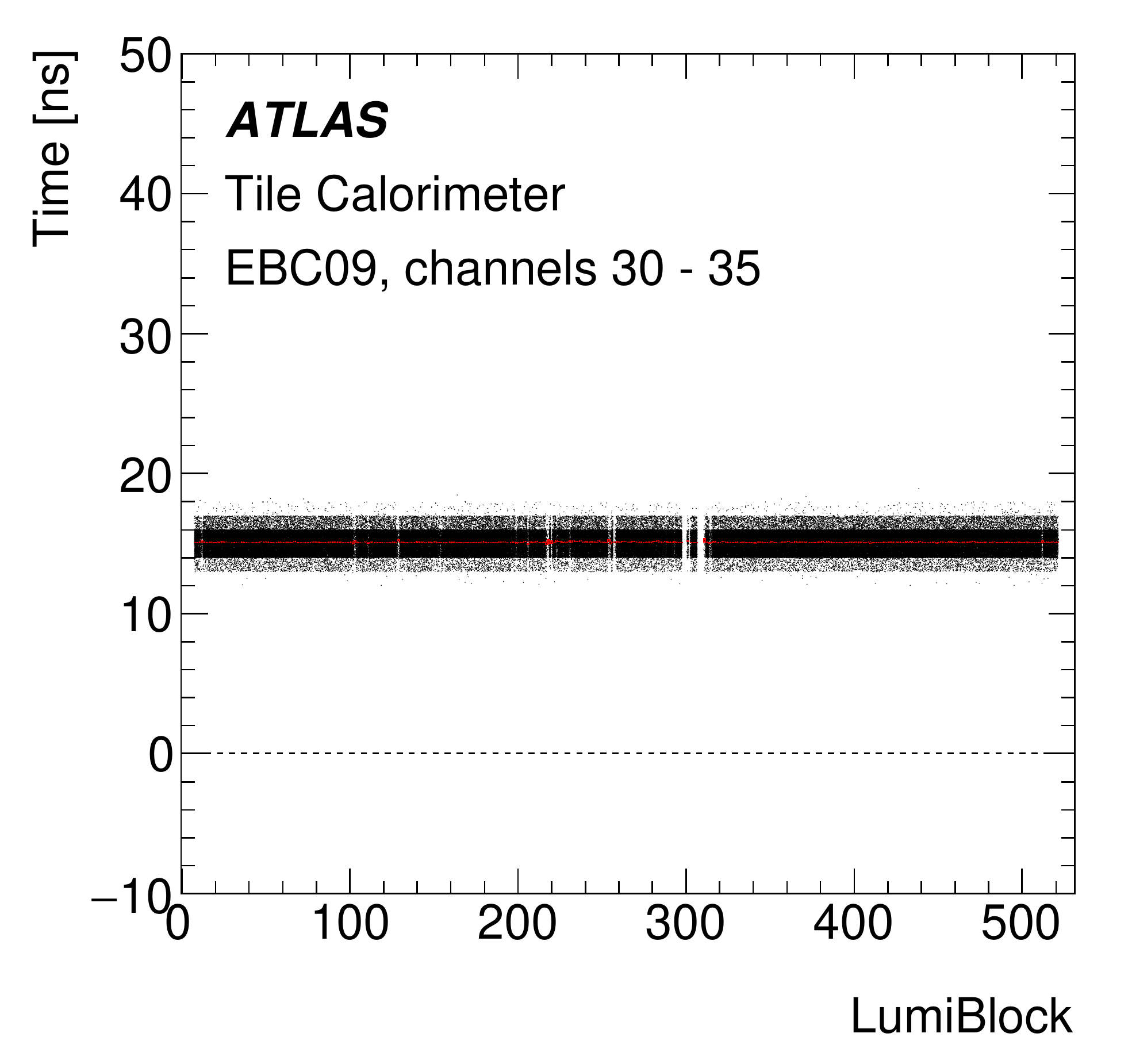}}
    \subfloat[\label{fig:timing_jump_b}]{\includegraphics[width=0.5\linewidth]{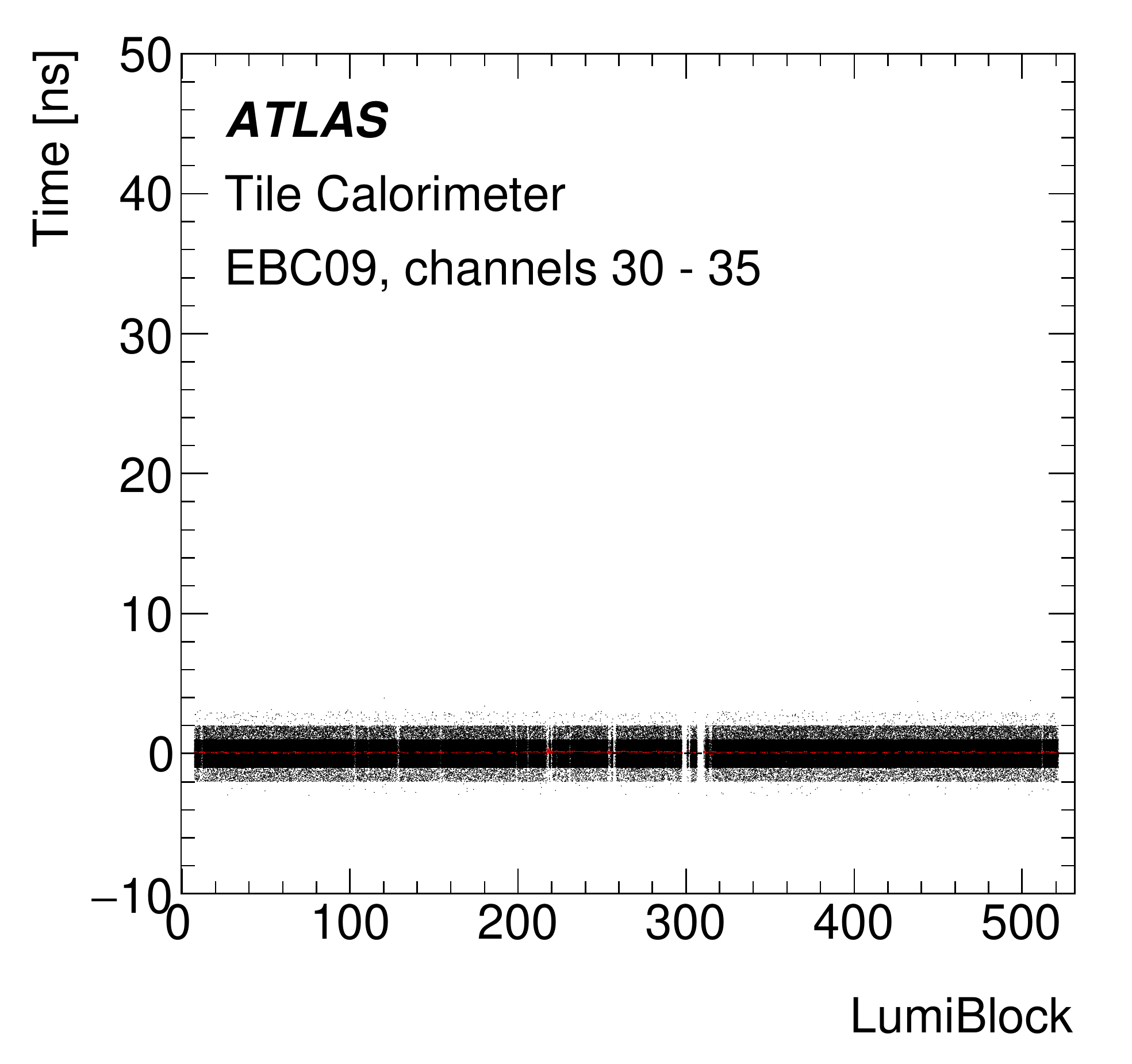}}
    \caption{An example of the timing jump of +15~ns in EBC09
    channels 30--35 before (a) and after (b) the time constant
    correction as identified with the laser-based monitoring tool. The
    dashed line indicates the expected mean time value.}
    \label{fig:timing_jump}
\end{figure}

The former feature typically indicates a sudden change of the timing settings of
the corresponding digitiser, so-called timing jump. These timing jumps
are corrected by adjusting the associated time constant in the affected
period, as shown in Figure~\ref{fig:timing_jump}.
While the timing jumps were very frequent during
Run~1~\cite{bib:laser_run_1} and a lot of effort was invested into their 
correction, they appeared very rarely during Run~2 due
to improved stability of the electronics. This allowed us to focus on
other problems observed with the monitoring tool.

\begin{figure}[t]
\centering
    \subfloat[\label{fig:time_BC_offset_a}]{\includegraphics[width=0.5\linewidth]{figures/laser_339037_eba40_final_500x500_1}}
    \subfloat[\label{fig:time_BC_offset_b}]{\includegraphics[width=0.5\linewidth]{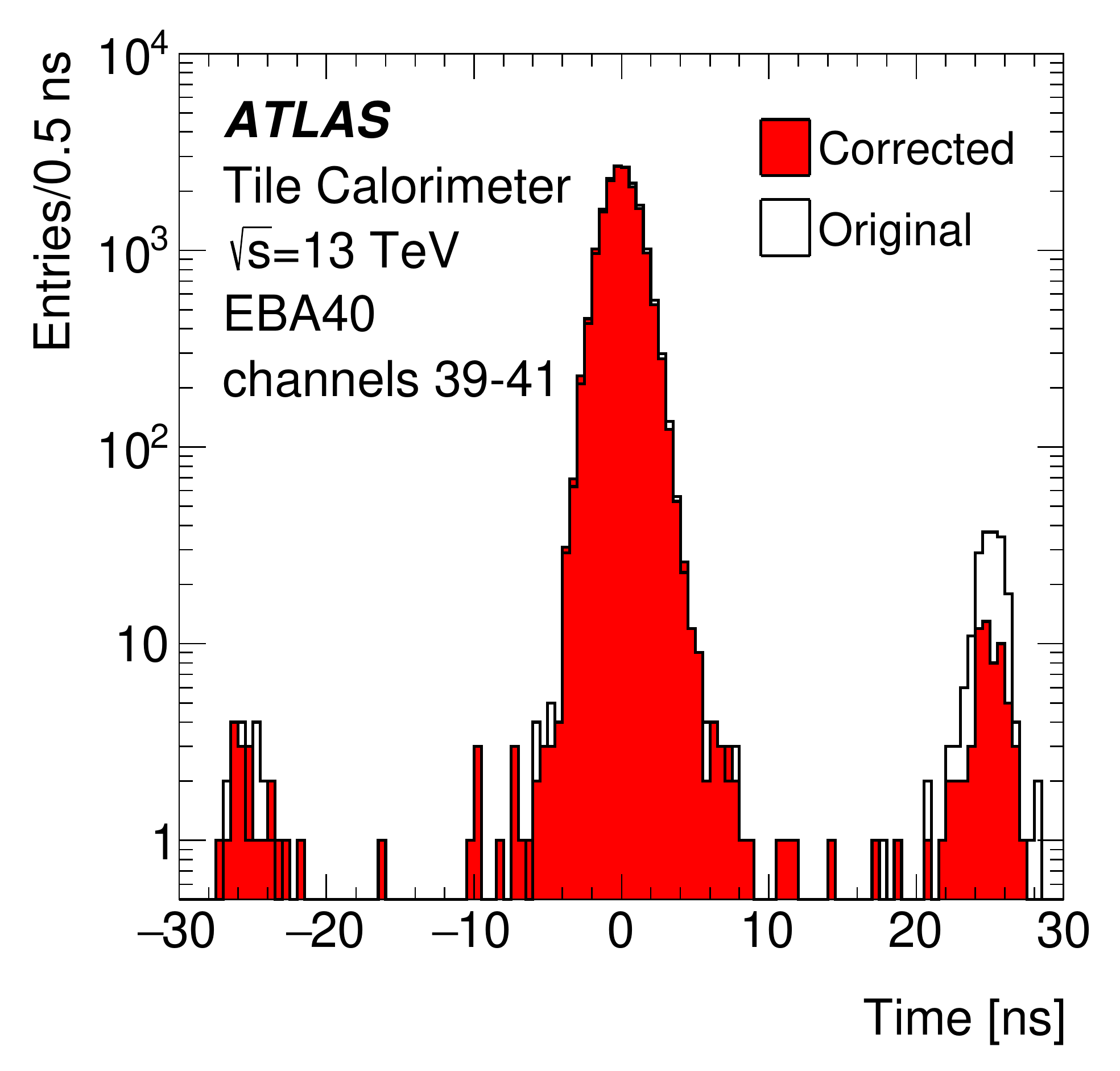}}
    \caption{The reconstructed time of laser events as a function of
    luminosity blocks (a): three channels belonging to the same
    TileDMU are superimposed. The majority of events, centred around zero, are well
    timed-in. The events with the bunch-crossing offset are centred at
    +25~ns and these events are fully correlated across the
    three channels. The reconstructed time in physics events in
    the same three channels before (original) and after (corrected)
    the algorithm mitigating the bunch-crossing offset events
    applied (b): the algorithm reduces by a factor 4 the events centred around
    +25~ns.}
    \label{fig:time_BC_offset}
\end{figure}

Few channels suffer from {\tlaser} sometimes off by 1 or 2
bunch-crossings, i.e. $\pm25$ or $\pm50$~ns. This feature
affects all three channels managed by the same Data Management Unit
(TileDMU)~\cite{Berglund:2008zz}. The problem is intermittent, with a 
rate at a percent-level; nevertheless, the observed
bunch-crossing offset and affected events are fully correlated across
the three channels. An example is shown in Fig.~\ref{fig:time_BC_offset_a}.
Such events also occur in physics collision data at a very similar
rate as in the laser data. Studies have shown that a difference of
25~ns between the actual and supposed time phases degrades the
reconstructed energy by 35\%~\cite{TCAL-2017-01}. For this reason, a dedicated software
tool was developed to detect cases affected by the bunch-crossing
offset in physics data on-the-fly and prevent them from propagation to
subsequent object reconstruction. Figure~\ref{fig:time_BC_offset_b} compares the reconstructed time in affected channels before and
after this tool is applied. The affected events close to +25~ns
are clearly reduced.  

\subsection{Dependence of the PMT response on the anode current}

The assumption of a linear relationship between the PMT signal amplitude and the cell energy deposits requires the PMT response to be independent of current. For most cells, the range of currents is small enough that any non-linearity is negligible. In contrast, highly exposed cells, such as the E~cells, experience a large current range between low and high luminosity runs, and between a caesium calibration run and a physics run. Therefore, those cells provide the necessary data to investigate such effects and are used to study the PMT response as a function of the anode current. Particular attention is paid to the difference between response of the E1 and E2~cells with respect to the E3 and E4~cells. The latter are the most exposed TileCal cells, where the larger particle fluence results in larger PMT currents. PMTs with active HV dividers~\cite{ATLAS-TDR-28} are installed in the readout of these cells to mitigate the current dependence, in principle affording larger stability. How well they do so must be understood when using the cells across a wide current range.

The measurement of the anode current comes from data of the TileCal readout of minimum-bias events that are collected during each run.
These minimum-bias data are read out via slow current integrators which were installed for the readout of low signals from the radioactive caesium source used in the calorimeter calibration. The integrators average the current in each cell over a long time window of 10--20~ms to suppress fluctuations in event-to-event energy deposition, diverting only a small fraction of each PMT's output from the primary signal. Large depositions from hard-scattering are also suppressed on long time scales.

Laser-in-gap data and the current measurements of minimum-bias events were analysed for three runs taken in 2018. The particular set of runs were selected to explore a wide current range while minimising the overlap of currents. The PMT signal amplitudes caused by the laser pulses are first pedestal-subtracted, primarily present due to electronics noise, which comes mostly from the front-end electronics used to shape the signal for the ADC, as well as from the presence of beam-induced and other non-collision background. The signal amplitude is not normalised to the reference diode, as done to determine the PMT calibration described in Section~\ref{sec:determination_of_the_calibration_constants}, since the small instability associated to this monitoring device is often larger than the effects being studied, and so are the uncertainties associated with its correction. Instead, a cleaner approach to minimise the impact of laser intensity fluctuations is adopted, normalising the measurements of the channels from E~cells of interest to a reference TileCal PMT with negligible current range on the same module. In this study, the left PMT of the D6~cell is used as the reference. The E~cell normalised response $R^{\mathrm{E/D6_L}}$ is defined per module as the ratio between the signal amplitudes of the E~cell PMT ($A^\mathrm{E}$) and D6 left PMT ($A^{\mathrm{D6_L}}$):

\begin{equation}\label{eq:Ecell_nD6L}
R^{\mathrm{E/D6_L}} = \frac{\mathrm{A^E}}{A^{\mathrm{D6_L}}}
\end{equation}

The minimum-bias current decays throughout a physics run as the proton beams decay, so the normalised E cell response changes as well if there is a dependence on the current. To determine this dependence, the actual cell response at any given current is compared to the nominal response at zero current. Therefore, the measurement in any given luminosity block is normalised to the mean measurement in the zero current period, i.e. before collisions begin:

\begin{equation}\label{eq:normalisedEcell_nD6L}
\frac{R^{\mathrm{E/D6_L}}}{R_{\mathrm{current=0}}^{\mathrm{E/D6_L}}}
\end{equation}

\begin{figure}[t]
\begin{center}
\subfloat[E1]{\includegraphics[height=0.49\textwidth,angle=-90, trim=6cm 0 0 0, clip=true]{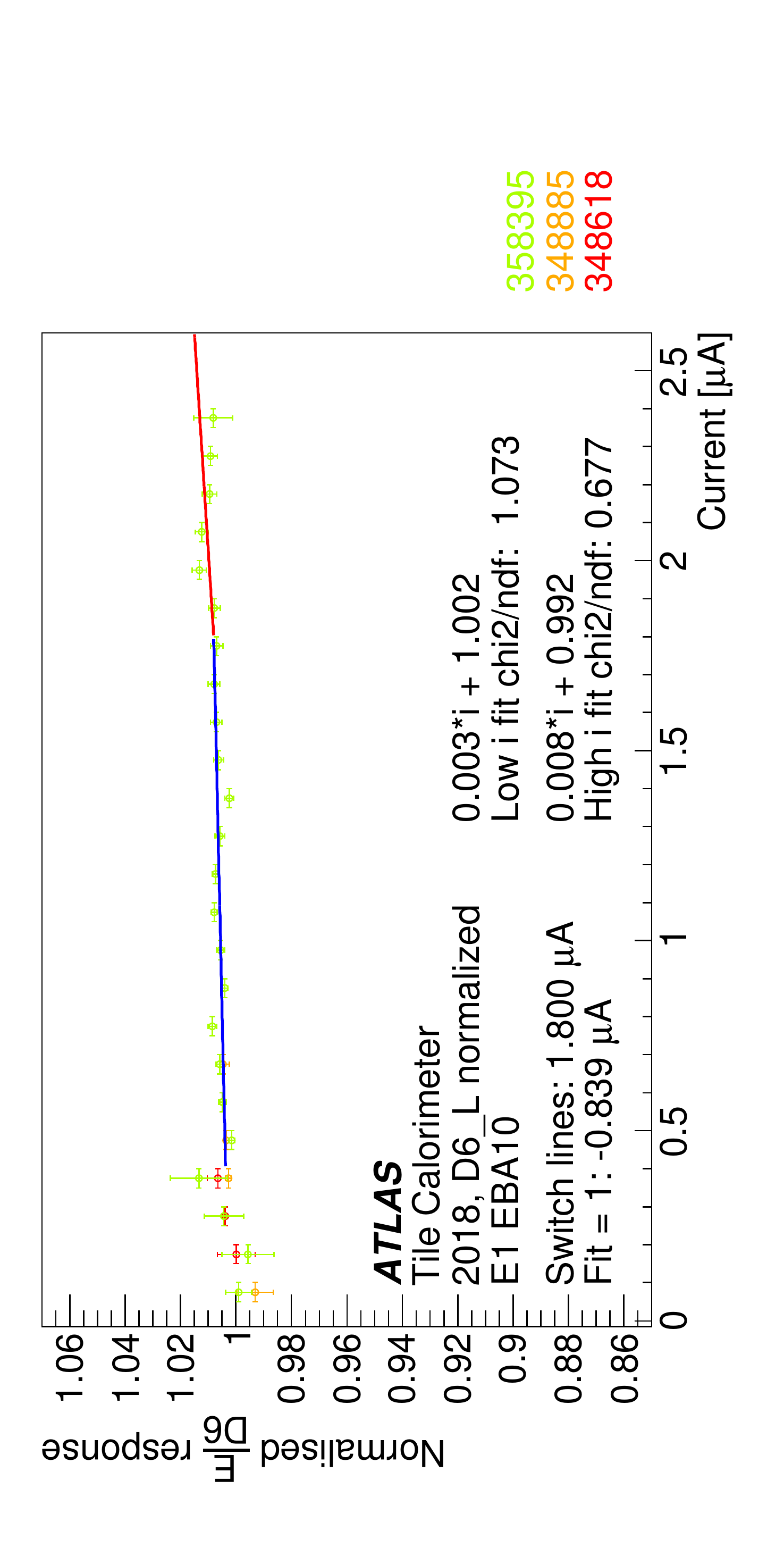}}
\subfloat[E2]{\includegraphics[height=0.49\textwidth,angle=-90, trim=6cm 0 0 0, clip=true]{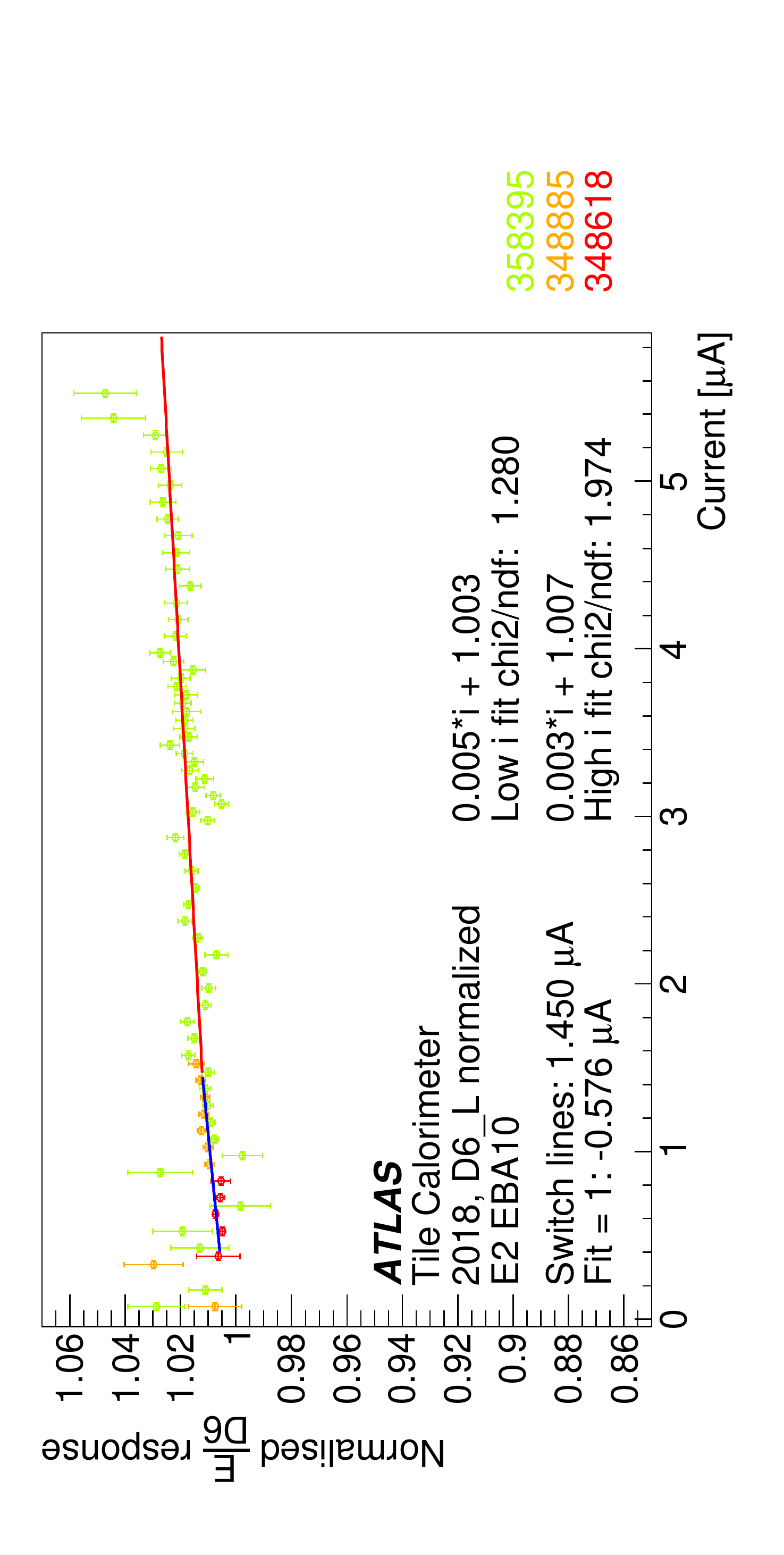}}\\
\subfloat[E3]{\includegraphics[height=0.49\textwidth,angle=-90, trim=6cm 0 0 0, clip=true]{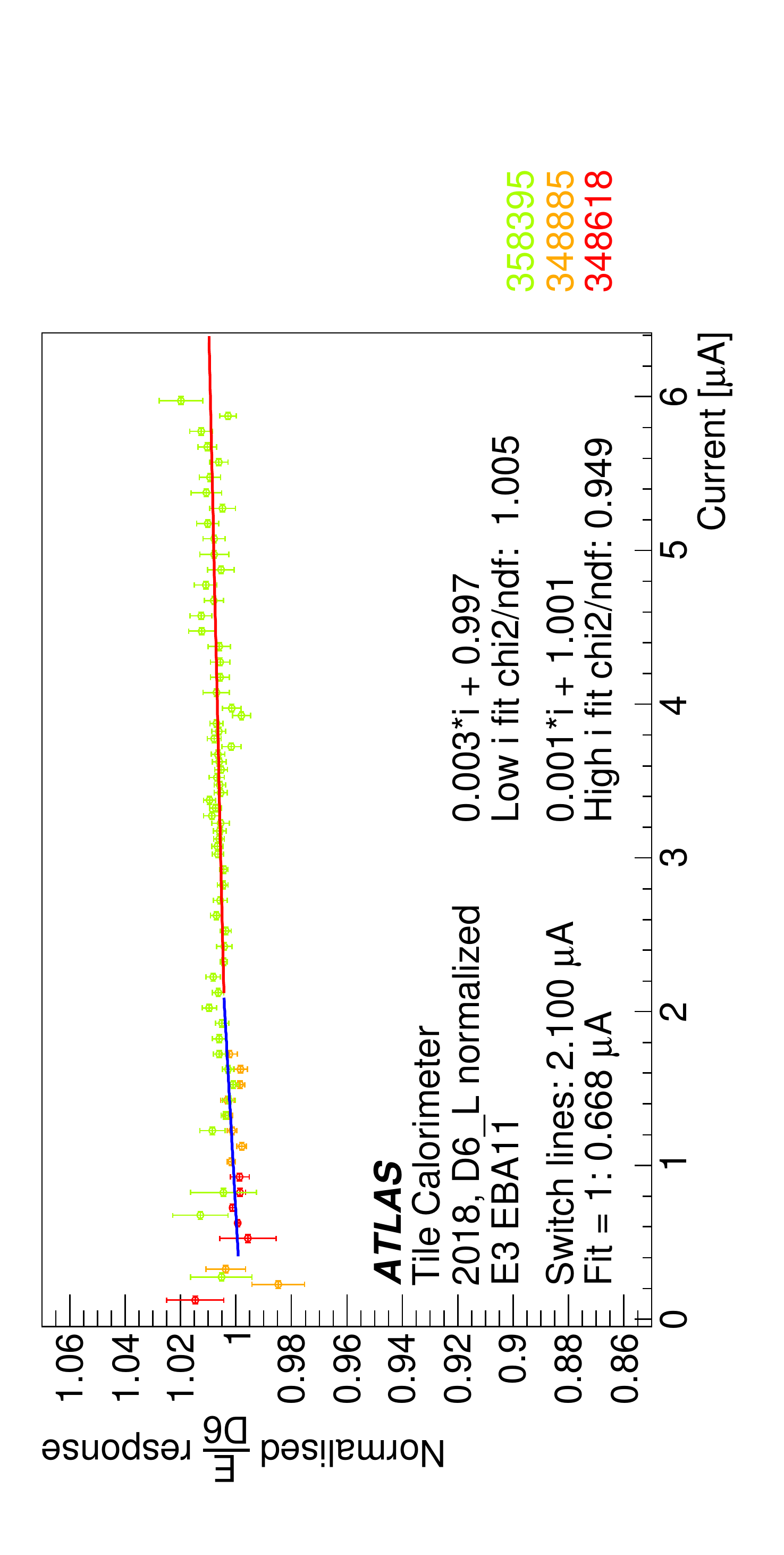}}
\subfloat[E4]{\includegraphics[height=0.49\textwidth,angle=-90, trim=6cm 0 0 0, clip=true]{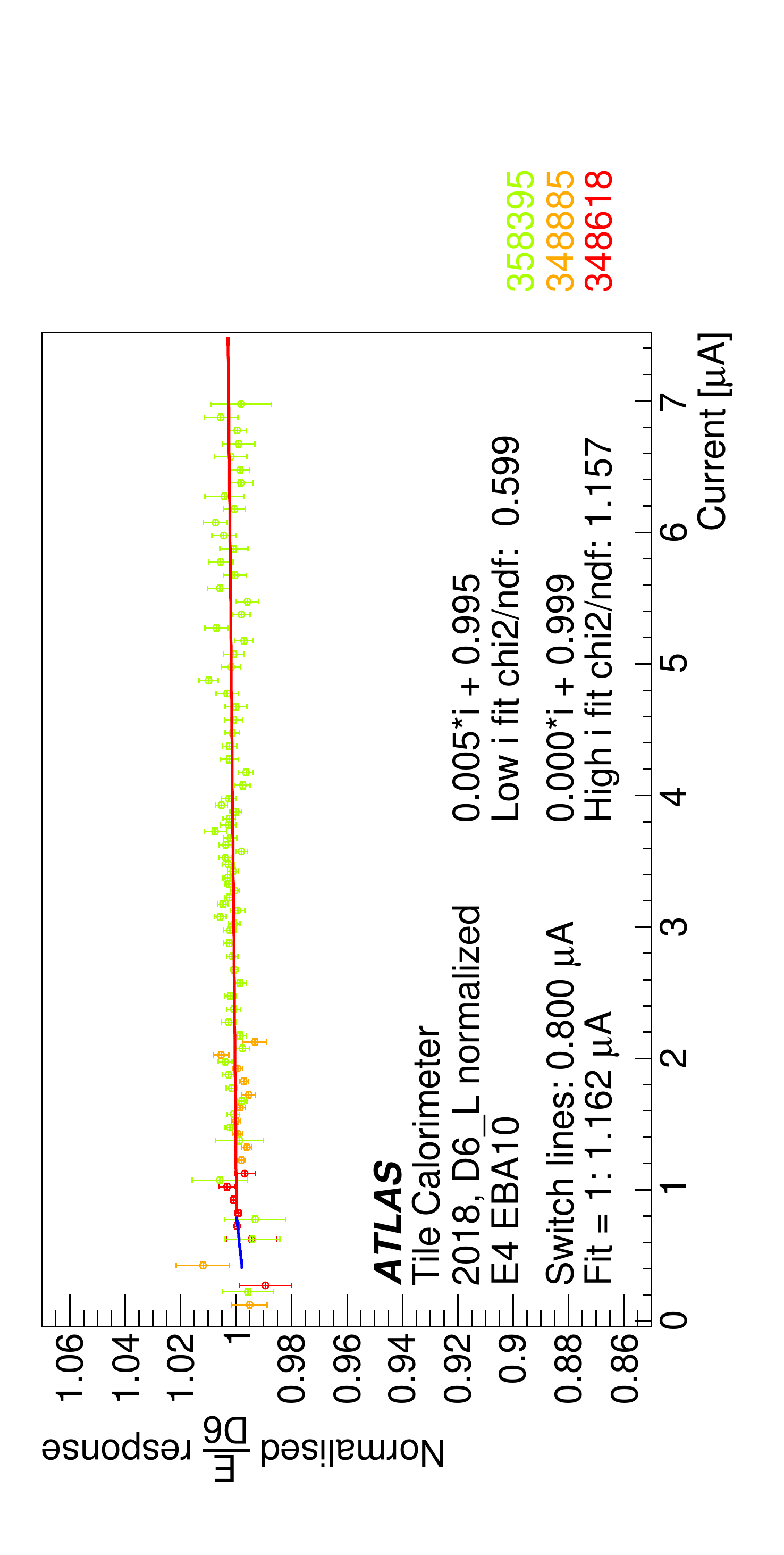}}
\caption{Normalised E/D6 cell response ratio as a function of current for an example channel from each family. The low current fit is in blue, while the high current fit is in red. ``Switch lines'' indicates the transition current between the low and high current fits. ``Fit = 1'' indicates the current at which the normalised E/D6 cell response ratio intercepts 1, with a negative value resulting in a non-physical intercept at 0 current. The fitted ratio can be applied as a function of current to correct for the current-dependence of the PMT response.}
\label{fig:fittedLiG}
\end{center}
\end{figure}

The baseline used for normalisation is the average E cell response ratio before stable beam declaration, where the first luminosity block with non-zero luminosity appears in the collision run. For each luminosity block of the chosen run, the average and RMS of the E/D6 cell response ratio to laser-in-gap pulses is calculated and normalised to the equivalent quantity at zero current. This is plotted as a function of the average anode current measured with the integrator readout of physics signals during the same luminosity block as shown in Figure~\ref{fig:fittedLiG} for the combination of the three selected runs.

Data are pruned by luminosity block if the laser pedestal is not stable or if the number of measurements in the luminosity block is less than 100. Luminosity blocks with a small number of measurements typically overlap with emittance scans or beam adjustments performed by ATLAS and in which the laser is disabled, so these are discarded. To further smooth the data, the measurements are averaged every $50~\mu\mathrm{A}$. The minimum current is chosen to avoid using data from fluctuations in the zero current measurement. The data are adjusted with a piecewise pair of linear fits:

\begin{itemize}
  \item A low current fit from $0.08~\mathrm{\mu A}$ until the transition current
  \item A high current fit from the transition current using all data up through $10~\mathrm{\mu A}$ 
\end{itemize}

The transition current is chosen by calculating the combined $\chi^2/n_{\mathrm{DoF}}$ for the two linear fits with possible transition currents in steps of $0.05~\mathrm{\mu A}$ up to a maximum possible transition current value of $2.30~\mathrm{\mu A}$. The current yielding the minimum $\chi^2/n_{\mathrm{DoF}}$ is chosen as the endpoint of the first linear fit and the beginning of the second fit, with piecewise continuity enforced. The procedure is also shown in Figure~\ref{fig:fittedLiG}. The transition current between low and high current regimes ranges between 0.8 and 2.1$\mathrm{\mu A}$. The PMT response dependence on current is stronger for E1 and E2 cells than for E3 and E4 cells where the active dividers were installed, especially in the high current regime. These results bring evidence that the active dividers are effectively stabilising the PMT response across a wide range of current operation.

Such a study can be used to determine a correction to calibrate the PMT response over current from the normalised PMT response ratio fitted function. In  Figure~\ref{fig:fittedLiG}, it can be seen that a maximum 2--3\% correction would be necessary at extreme high current for E1 and E2 cells, respectively. This requires a precise current measurement throughout the range of currents that may be experienced by the different cells. This study demonstrates that such correction should be more important for cells with passive dividers experiencing higher currents, including A cells. 


\section{Channel monitoring and PMT linearity}
\label{sec:monitoring}

\subsection{Automated channel monitoring}

Channels having pathological problems need to be promptly identified during the data taking periods. Therefore, an automated daily monitoring utilising the laser system is setup in order to identify and diagnose the channels' issues. This is achieved by analysing the recent laser calibration runs. After each laser run, several monitoring figures are produced by an automated software in order to control the PMT stability and provide the list of problematic channels with possible source of issues. All TileCal channels, including the masked channels after data quality checks, are analysed and flagged according to the algorithm described below.

The automated laser monitoring algorithm is based on the analysis of the PMT response variation, measured for each channel using the Direct method (explained in Section~\ref{sec:determination_of_the_calibration_constants}), and the comparison with other data (applied HV, global behaviour of group of channels associated to the same cell type). 

The time evolution of the channels' response is daily monitored using LG and HG laser runs taken in 15 preceding days, and three categories of channels are defined:

\begin{itemize}
  \item \textbf{Normal channels:} Channels that have no deviation or a deviation compatible with the mean deviation of similar cells. These channels can be calibrated safely and do not require a special attention.
  \item \textbf{Suspicious channels:} Channels with a deviation slightly higher than the mean deviation of similar cells or a deviation compatible with the one expected from the variation of the HV supply~\footnote{Taking into account that the PMT gain scales with $\mathrm{HV}^\beta$, where the $\beta$ parameter depends on the photomultiplier model. The nominal $\beta$ value for the TileCal PMTs is 6.9.}. These channels can be calibrated safely but some follow up may be needed.
  \item \textbf{Channels to be checked:} Channels with large deviations (>10\%) that cannot be explained by the mean deviation observed in cells of similar type nor by HV changes; or channels having a non linear behaviour during the 15 preceding days (jumps or fast drifts). These channels should not be calibrated unless the origin of the effect is understood. In most of the cases, especially in case of a fast drift, the channels need to be masked.
\end{itemize}


While this categorisation assists during channel calibration, the automated monitoring of laser data also identifies channels with pathological behaviour and determines the source of the issues encountered, complementing data quality assessment activities. During the data taking period, laser runs are chosen with approximately 10 days interval. For each chosen date, both HG and LG runs are analysed and channels are classified according to their problems as follows:

\begin{itemize}
  \item \textbf{Bad channels:} Channels with large PMT drift (>10\%) or having a wrong behaviour during the 15 preceding days.
  \item \textbf{HV unstable:} Channels with large PMT drift (>10\%) but compatible with HV variation.
  \item \textbf{No laser data:} Channels with low laser signal amplitude (e.g. caused by a laser fibre problem).
  \item \textbf{Bad laser data:} Channels with corrupted laser calibration data or having problematic data in the reference run.
\end{itemize}

\begin{figure}[t]
  \centering
  \subfloat[\label{fig:overall_a}]{\includegraphics[height=0.35\linewidth]{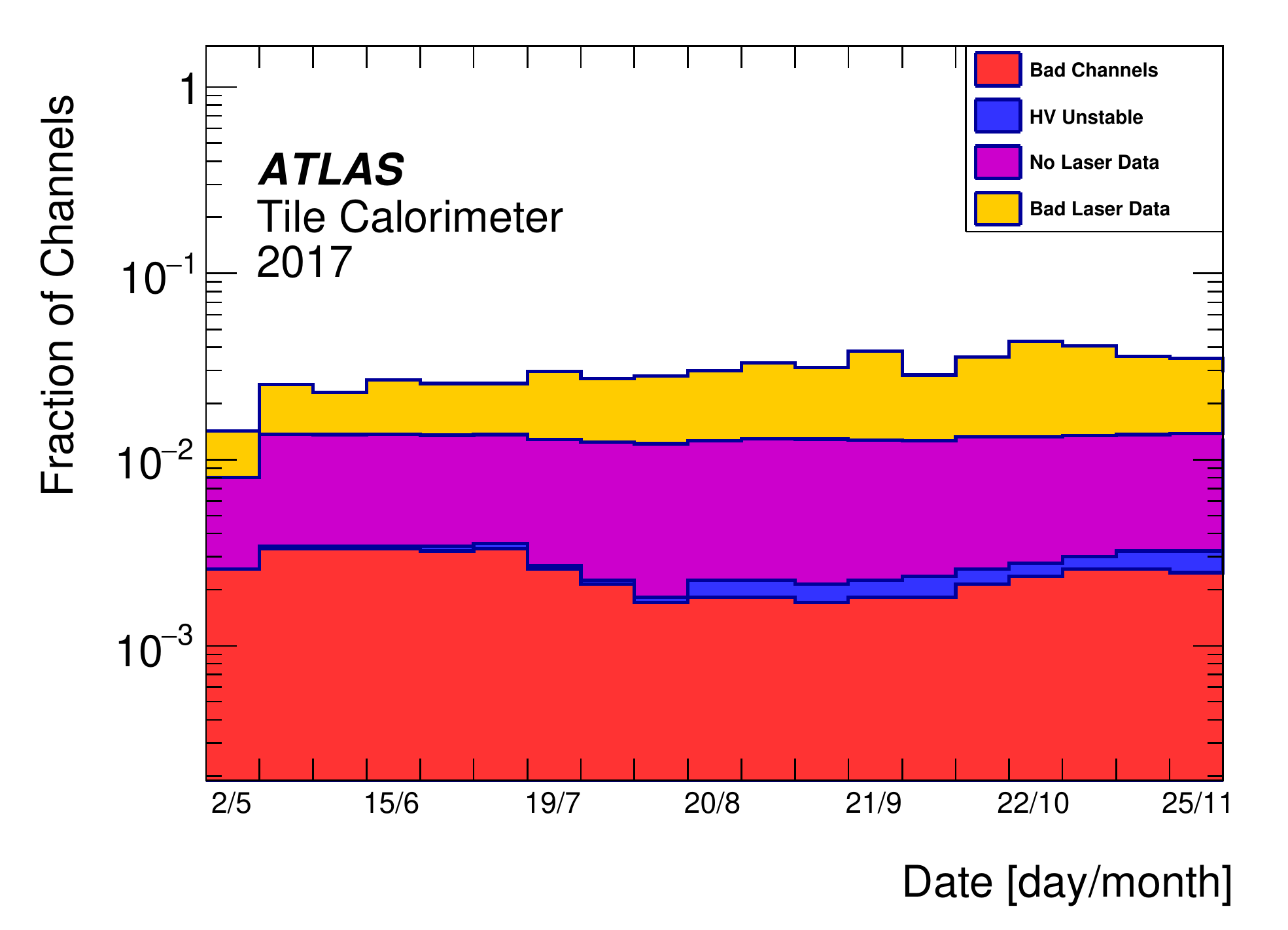}}\quad
  \subfloat[\label{fig:overall_b}]{\includegraphics[height=0.35\linewidth]{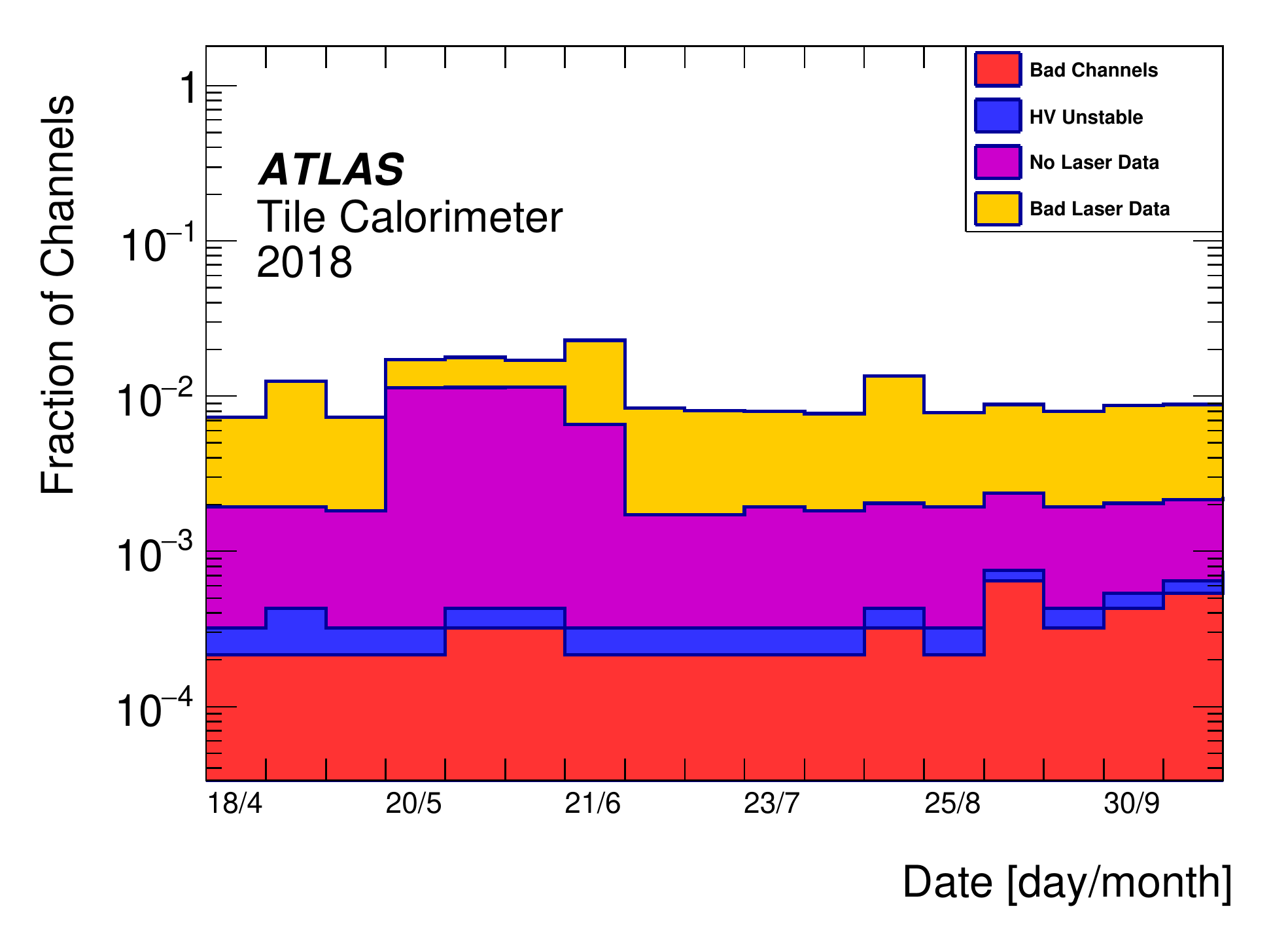}}
    \caption{\label{fig:overall} The fraction of the problematic channels identified by the  laser monitoring algorithm in 2017 (a) and 2018 (b).
    } 
\end{figure}

A channel is reported to be problematic if it manifests any type of issue listed above. Figure~\ref{fig:overall} shows the fraction of problematic channels, observed in 2017 and 2018, as a function of time. The maximum number of such channels did not exceed 4 and 3\%, respectively.

\FloatBarrier

\subsection{PMT linearity monitoring}

The TileCal PMT calibration is performed using laser light with a constant intensity. This procedure contributes to ensure that the calorimeter measures the same output over time for the same input energy deposition, i.e. that its response is stable. However, to guarantee that the calibration factors are accurate across the entire dynamic range of the PMT response and the output signal is directly proportional to the energy deposit one needs to assess the PMT linearity.

The linearity of the TileCal PMT channels was monitored during the Run~2 operation with laser calibration data acquired between 2016 and 2019. The dataset corresponded to a combination of standard laser calibration low gain runs using different filter wheel positions and laser intensity varied in the range of 12k to 18k in DAC counts. The linearity of a given PMT channel is evaluated by comparing the PMT signal to the signal of the reference photodiode D6 of the Laser~II system. The response of the channels should increase linearly with the light intensity, in the same way that the PMT channels respond to increases in the energy deposited in the calorimeter. 

PMTs lose linearity shortly before the saturation point. In addition, the TileCal ADCs saturate at an upper limit of 1023 counts. The saturation amplitude is given by $A_{\mathrm{max}}\;\mathrm{[pC]} = (1023-p)/f_\mathrm{ADC\to pC}$, where $p$ and $f_\mathrm{ADC\to pC}$ are the pedestal and CIS constant values, respectively. Values above the ADC saturation amplitude are not reliable and any non-linear behaviour above it should not be related exclusively to the PMT. Typically, TileCal readout channels start to loose linearity above $\sim750$~pC and reach saturation at $\sim850$~pC. This behaviour can be observed for the TileCal channels that receive enough light. To avoid this issue, amplitudes above $A_{\mathrm{max}} - \sigma_A$, where $\sigma_A$ is the standard deviation of the amplitudes, are excluded from the analysis.

\begin{figure}[t]
\centering
\subfloat[\label{fig:PMT_fit}]{\includegraphics[width=0.45\textwidth]{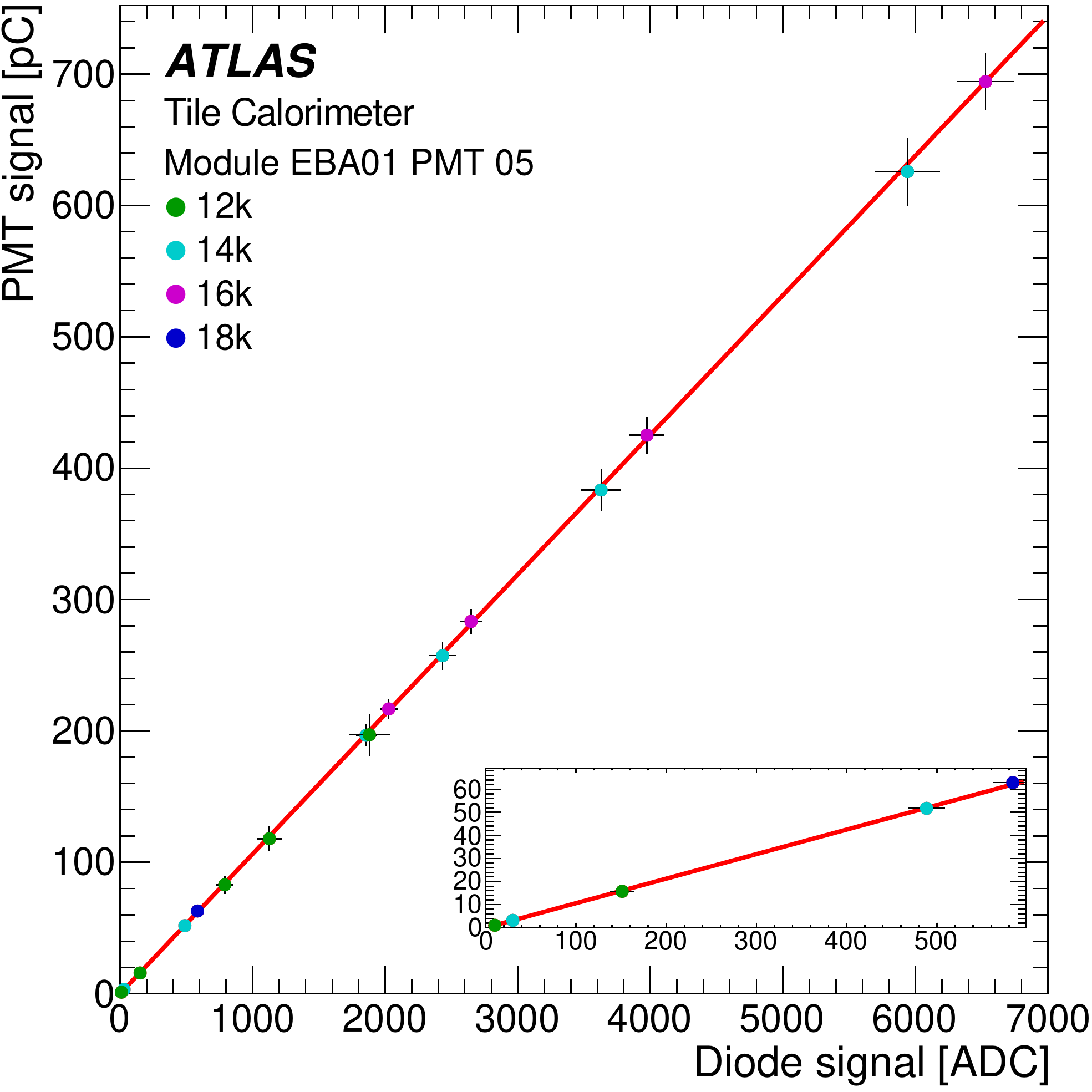}}\hfill
\subfloat[\label{fig:integral}]{\includegraphics[width=0.48\textwidth]{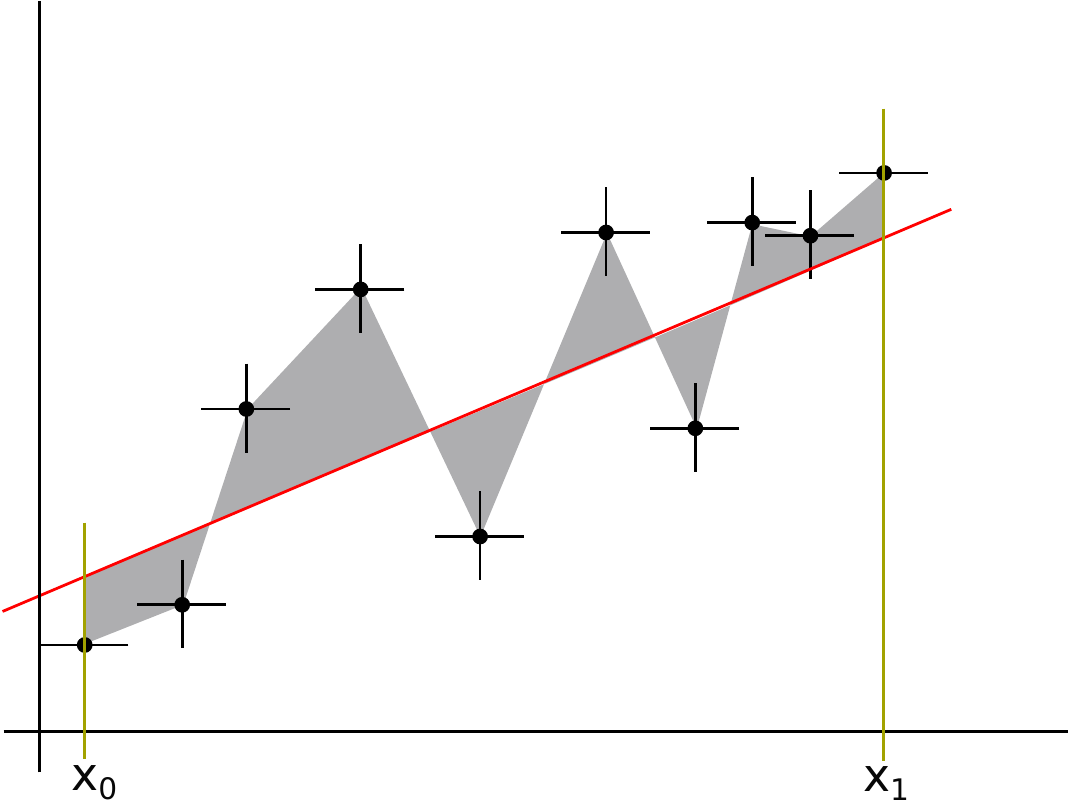}}
\caption{(a) Signal in EBA01 PMT 5 (channel 4) versus signal in photodiode~6 using the dataset taken on 2016-07-10. The red line shows the obtained fit. The inset shows the magnified distribution of the low laser-intensity region. (b) Schematic representation of the area obtained by intercepting the joint data points and the fit function, in grey. The deviation from linearity is defined as the ratio between the grey area and the fit function integral between the first ($x_0$) and last ($x_1$) points.}
\end{figure}

\begin{figure}[t]
\centering
\includegraphics[width=0.5\textwidth]{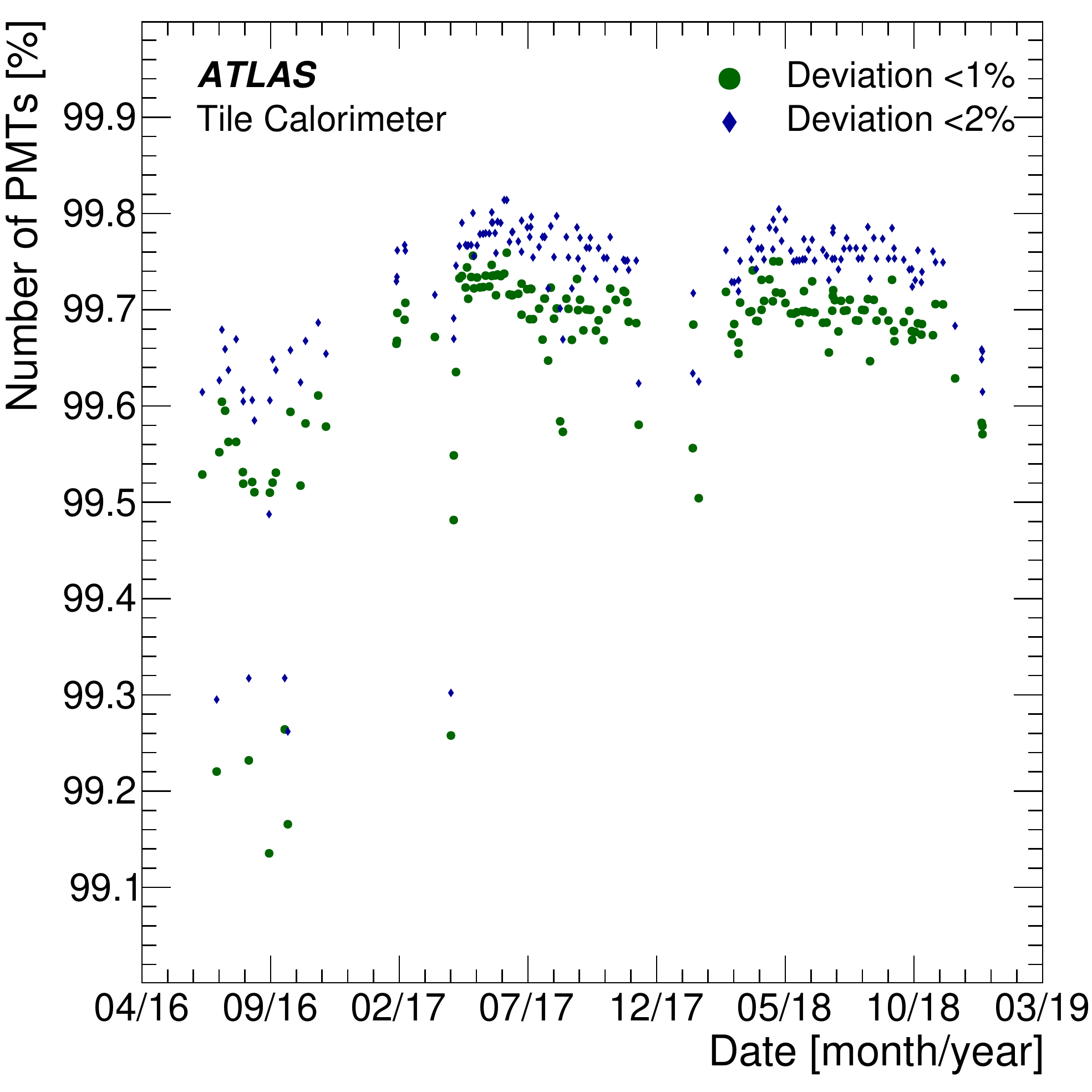}
\caption{The percentage of PMTs within 1\,\% and 2\,\% deviation from linearity as a function of time for the weekly calibration runs taken between 2016-07-10 and 2019-01-18 are shown.\label{fig:linearity}}
\end{figure}

The PMT signal amplitude versus the reference diode signal is plotted for a set of runs of varied light output taken in the same day. A linear fit is performed iteratively for each data point and all the points with smaller amplitudes. The fit comprising more points within one standard deviation of the fitted line is chosen as final for further analysis. 
An example can be seen in Figure~\ref{fig:PMT_fit}.

The deviation from linearity, in percentage, is defined as the ratio between the area delimited by data points intercepted by the linear fit, and the integral of the linear fit, as shown in Figure~\ref{fig:integral}. 

Figure~\ref{fig:linearity} shows the percentage of PMTs within 1~\% and 2~\% deviation from linearity as a function of time for the weekly calibration runs taken between 2016 and 2019. The percentage of channels with deviation from linearity less than 1\% is $(99.66 \pm 0.11)\%$ and less than 2\% is $(99.72 \pm 0.09)\%$ considering this period.

\FloatBarrier

\FloatBarrier

\section{Conclusion}
\label{sec:conclusion}

The Laser~II calibration system of the ATLAS Tile Calorimeter probes individually the 9852 PMTs of the detector, together with the readout electronics of each channel. It is one of the three dedicated systems ensuring the calibration of the full calorimeter response. The Laser~II system has undergone a substantial upgrade during the LHC Long Shutdown 1 that improved its stability for the calibration of the calorimeter in Run~2. The readout electronics was renewed and a new light splitter was installed; the system now includes ten photodiodes to monitor the light across its path in the Optics box.

Laser runs were used to regularly determine the PMT response and calculate the calibration constants $f_\mathrm{Las}$, which were updated weekly for the cell energy reconstruction. The PMT response fluctuations are highly correlated with the LHC operations, with the response decreasing with integrated current and recovering in following technical stops of the collider. The calibrations obtained with the laser and caesium source are consistent within 0.4\%, dominating the systematic uncertainty on the PMT calibration scale, improving over the laser systematic error found in Run~1. In addition, a sub-dominant uncertainty on the PMT relative inter-calibration was found to have a luminosity dependence. The linearity of the PMTs was studied to ensure that the calibration factors are accurate across the dynamic range of the PMT response.


Laser events were also used to evaluate the timing of the readout electronics, monitor the stability of time calibration and detect pathological behaviours in the calorimeter channels in data quality activities, contributing to the high TileCal performance in Run 2 and to achieve the design goals of the experiment with respect to jet energy resolution of 3.5\% for central jets of very high transverse momenta~\cite{ATLAS:2020cli}, matching the design goals of the experiment.
Laser events were also used to evaluate the timing of the readout electronics, monitor the stability of time calibration and detect pathological behaviours in the calorimeter channels in data quality activities, contributing to the high TileCal performance in Run 2 and to achieve the design goals of the experiment with respect to jet energy resolution.

\section*{Acknowledgments}

The authors would like to acknowledge the entire TileCal community for their contribution with the discussions related to this work, the operations acquiring laser calibration data, the input from the data quality activities, and the careful review of this report.


\bibliographystyle{JHEP}
\bibliography{laserRun2Paper.bib}


\end{document}